\documentclass[11pt]{article}
\usepackage[latin1]{inputenc}
\usepackage{xspace}
\usepackage[english]{babel}
\usepackage{amsmath,amssymb,amsfonts,epic}
\usepackage[dvips]{graphicx}
\usepackage{graphics,epsfig}
\usepackage{fancybox}
\usepackage{amsbsy}
\usepackage{amscd}
\usepackage{amsgen}
\usepackage{amsopn}
\usepackage{amstext}
\usepackage{ifthen}
\usepackage{amsxtra}
\usepackage{color}
\usepackage{fancyhdr}
\usepackage{fullpage}
\usepackage{float}
\usepackage{arydshln}
\usepackage[round]{natbib}
\definecolor{hellgrau}{gray}{0.4}

\def \Lam{\Lambda(\lambda,J_0^c)}
\def\argmin{\mathrm{argmin}}

\newcommand{\normu}[1]{\ensuremath{\vert\!\vert #1 \vert\!\vert}_{\ell_1}}
\newcommand{\normd}[1]{\ensuremath{\vert\!\vert #1 \vert\!\vert}_{\ell_2}}

\newcommand{\normD}[1]{\ensuremath{\vert\!\vert #1 \vert\!\vert}_{n}}
\newcommand{\E}{\ensuremath{\mathbb{E}}}

\renewcommand{\P}{\ensuremath{\mathbb{P}}}

\newcommand{\R}{\ensuremath{\mathbb{R}}}

\newcommand{\la}{\ensuremath{\lambda}}

\newcommand{\e}{\ensuremath{\varepsilon}}
\newcommand{\p}{\ensuremath{\varphi}}

\newcommand{\Ne}{\ensuremath{\mathbb{N}}}
\newcommand{\fo}{f}
\newcommand{\Jo}{\ensuremath{J_0}}

\newcommand{\Comp}[1]{\ensuremath{#1^C}}
\newcommand{\JoC}{\ensuremath{\Comp{\Jo}}}

\newcommand{\betamod}{\alpha}

\newtheorem{lemma}{Lemma}
\newtheorem{Th}{Theorem}
\newtheorem{cor}{Corollary}
\newtheorem{ex}{Example}
\newtheorem{assump}{Assumption}
\newcommand{\bdGamma}{\boldsymbol\Gamma}
\newcommand{\bdeta}{\boldsymbol\eta}
\newcommand{\bdbeta}{\boldsymbol\beta}
\newcommand{\bdphi}{\boldsymbol\phi}
\newcommand{\bdy}{\boldsymbol y}
\newcommand{\bdI}{\mbox{\bf I}}
\newcommand{\bdg}{\boldsymbol g}
\newcommand{\bdf}{\boldsymbol f}
\newcommand{\bdx}{\boldsymbol x}
\newcommand{\bdz}{\boldsymbol z}
\newcommand{\bdA}{\boldsymbol A}

\newcommand{\bdtheta}{\boldsymbol\theta}

\newcommand{\diag}{\mbox{diag}}

\begin{document}

\title{Lasso-type estimators for Semiparametric Nonlinear
Mixed-Effects Models Estimation}
\author{Ana Arribas-Gil
\\Departamento de Estad\'istica\\
        Universidad Carlos III de Madrid, Getafe, Spain.\\ E-mail: aarribas@est-econ.uc3m.es
\and Karine Bertin
\\CIMFAV-Facultad de Ingenier\'ia\\
        Universidad de Valpara\'iso, Valpara\'iso, Chile.\\ E-mail: karine.bertib@uv.cl
\and Cristian Meza
\\CIMFAV-Facultad de Ingenier\'ia\\
        Universidad de Valpara\'iso, Valpara\'iso, Chile.\\ E-mail: cristian.meza@uv.cl
\and Vincent Rivoirard
\\CEREMADE, CNRS-UMR 7534, Universit\'e Paris Dauphine, Paris, France.\\ E-mail:Vincent.Rivoirard@dauphine.fr
}
\date{}
\maketitle

\thispagestyle{empty}

\begin{abstract}
Parametric nonlinear mixed effects models (NLMEs) are now widely used in biometrical studies, especially
in pharmacokinetics research and HIV dynamics models, due to, among other aspects, the computational advances
achieved during the last years. However, this kind of models may not be flexible enough for complex longitudinal
data analysis. Semiparametric NLMEs (SNMMs) have been proposed by Ke and Wang (2001).
These models are a good compromise and retain nice features of both parametric and nonparametric models resulting
in more flexible models than standard parametric NLMEs. However, SNMMs are complex models for which estimation still remains a challenge. The estimation procedure proposed by \cite{KeWang} is based on a combination of log-likelihood approximation methods for parametric estimation and
smoothing splines techniques for nonparametric estimation. In this work, we propose new
estimation strategies in SNMMs. On the one hand, we use the Stochastic Approximation version of EM algorithm (Delyon et al., 1999) to
obtain exact ML and REML estimates of the fixed effects and variance components. On the other hand, we propose a LASSO-type method to estimate the unknown nonlinear function. We derive oracle inequalities for this nonparametric estimator.
We combine the two approaches in a general estimation procedure that we illustrate with simulated and real data.
\end{abstract}



\section{Introduction}
We consider the semiparametric nonlinear mixed effects model (SNMM) as defined by \cite{KeWang} in which we have $n$ individuals and we observe:
\begin{eqnarray}\label{modmixte}
y_{ij}&=& g(\bdx_{ij}, \bdphi_i, f) +\varepsilon_{ij}, \quad \varepsilon_{ij}\sim {\cal N}(0, \sigma^2) \mbox{ i.i.d.},
\quad i=1\dots,N, \,\, j=1,\dots,n_i \end{eqnarray}
where $y_{ij}\in\R$ is the $j$th observation in the $i$th individual, $\bdx_{ij}\in\R^d$ is a known regression
variable,  $g$ is a common known function governing within-individual behaviour and $f$ is an unknown nonlinear function to estimate. The random effects $\bdphi_i\in \R^p$ satisfy
\begin{eqnarray*}\label{model2}
\bdphi_i&=& \boldmath{\bdA_i \bdbeta + \bdeta_i}, \quad \bdeta_i \sim {\cal N}(0, \bdGamma) \mbox{ i.i.d.}
\end{eqnarray*}
where $\bdA_i\in\mathcal{M}_{p,q}$ are known design matrices, $\bdbeta\in\R^q$ is the unknown vector of fixed effects and we suppose that $\varepsilon_{ij}$ and $\bdeta_i$ are mutually independent. We use bold letters for vector and matrices.\\
The parameter of the model is $(\bdtheta, f)$, where $\bdtheta=(\bdbeta, \bdGamma, \sigma^2)$ belongs to a finite
dimensional space whereas $f$ belongs to an infinite dimensional space of functions denoted ${\cal H}$.\\

\cite{KeWang} consider the most common type of SNMM in practice, in which $g$ is linear in $f$ conditionally to $\bdphi_i$,
\begin{equation}\label{formef}
g(\bdx_{ij}, \bdphi_i, f) = a(\bdphi_i; \bdx_{ij}) + b(\bdphi_i; \bdx_{ij})f(c(\bdphi_i; \bdx_{ij})),
\end{equation}
where $a$, $b$ and $c$ are known functions which may depend on $i$.\\

Different formulations of SNMM's have been recently used to model circadian rhythms (\cite{WangBrown}, \cite{WangKeBrown}), HIV dynamics (\cite{WuZhang}, \cite{LiuWu07}, \cite{LiuWu08}) or gene expression data (\cite{LuanLi}) among other applications.\\
\begin{ex}\label{humancircadian}
The following model was proposed by \cite{WangBrown} to fit human circadian rhythms:
\begin{eqnarray*}
y_{ij}=\mu+ \eta_{1i}+\exp(\eta_{2i})\,f\left(x_{ij} - \dfrac{\exp(\eta_{3i})}{1+\exp(\eta_{3i})}\right)+\varepsilon_{ij}, &&\varepsilon_{ij}\sim {\cal N}(0, \sigma^2) \mbox{ i.i.d.}\\ && \boldmath{\bdeta_i} \sim  {\cal N}(0, \bdGamma) \mbox{ i.i.d.}
\end{eqnarray*}
for $i=1\dots,N$, $j=1,\dots,n_i$, where $y_{ij}$ is the physiological response of individual $i$th at the $j$th time point $x_{ij}$. This model can be written in the general form (\ref{modmixte}) as:
\begin{eqnarray*}
y_{ij}&=& g(x_{ij}, \bdphi_i, f) +\varepsilon_{ij}, \quad \varepsilon_{ij}\sim {\cal N}(0, \sigma^2) \mbox{ i.i.d.}, \quad i=1\dots,N,\,\, \quad j=1,\dots,n_i \\
g(x_{ij}, \bdphi_i, f) &=& \phi_{1i} + \exp(\phi_{2i})\,f\left(x_{ij} - \dfrac{\exp(\phi_{3i})}{1+\exp(\phi_{3i})}\right)\\
\bdphi_i&=& (1,0,0)^{T} \mu+ \bdeta_i , \quad \bdeta_i \sim {\cal N}(0, \bdGamma) \mbox{ i.i.d.}
\end{eqnarray*}
where $\bdphi_i=(\phi_{1i},\phi_{2i},\phi_{3i})^{T}$ and $\bdeta_i=(\eta_{1i},\eta_{2i},\eta_{3i})^{T}$. In  this example $f$ represents the common shape of the observed curves, and $\phi_{1i}$, $\exp(\phi_{2i})$, and $\exp(\phi_{3i})/(1+\exp(\phi_{3i}))$ stand for the individual vertical shift, individual amplitude and individual horizontal shift respectively. Here $d=1$, $p=3$, $q=1$ and the parameter of the model is $(\mu,\bdGamma, \sigma^2, f)$. This model was also used by \cite{KeWang} for modeling Canadian temperatures at different weather stations.
\end{ex}

Let us introduce the following vectorial notations: $\bdy_i=(y_{i1},\dots,y_{in_i})'$, $\bdy=(\bdy'_1,\dots,\bdy'_N)'$, $\bdphi=(\phi'_1,\dots,\phi'_N)'$, $\bdeta=(\bdeta'_1,\dots,\bdeta'_N)'$, $\bdg_i(\bdphi_i,f)=(g(\bdx_{i1}, \bdphi_i, f),\dots,g(\bdx_{in_i}, \bdphi_i, f))'$, $\bdg(\bdphi,f)=(\bdg_1(\bdphi_1,f)',\dots, \bdg_N(\bdphi_n,f)')'$, $\bdA=(\bdA'_1,\dots,\bdA'_N)'$, $\widetilde{\bdGamma}=\diag(\bdGamma,\dots,\bdGamma)$ and $n=\sum_{i=1}^N n_i$. Then, model (\ref{modmixte}) can be written as:
\begin{eqnarray}\label{mod_vec}
\bdy|\bdphi&\sim& {\cal N}(\bdg(\bdphi,f),\sigma^2 \bdI_n)\nonumber\\
\bdphi&\sim& {\cal N}(\bdA \bdbeta,\widetilde{\bdGamma})
\end{eqnarray}
where $\bdI_n$ represents the identity matrix of dimension $n$.\\

The likelihood of observations $\bdy$ is:
\begin{eqnarray}\label{lik}
&&p(\bdy; (\bdtheta,f))= \int p(\bdy|\bdphi; (\bdtheta,f)) p(\bdphi; (\bdtheta,f)) d\bdphi \nonumber\\
&=&\int \dfrac{1}{(2 \pi \sigma^2)^{\frac{n}{2}}} \exp\left\{\dfrac{-1}{2\sigma^2} \|\bdy-\bdg(\bdphi,f)\|^2\right\}
\dfrac{1}{(2 \pi)^{\frac{Np}{2}} |\bdGamma|^{\frac{N}{2}}} \exp\left\{\dfrac{-1}{2}\|\widetilde{\bdGamma}^{-1/2}(\bdphi-\bdA \bdbeta)\|^2\right\} d\bdphi \nonumber\\
&=&\frac{1}{(2 \pi)^{\frac{n+Np}{2}} (\sigma^2)^{\frac{n}{2}} |\bdGamma|^{\frac{N}{2}}}\int \exp\left\{\dfrac{-1}{2} \left(\frac{1}{\sigma^2} \|\bdy-\bdg(\bdphi,f)\|^2 + \|\widetilde{\bdGamma}^{-1/2}(\bdphi-\bdA \bdbeta)\|^2\right)\right\} d\bdphi,
\end{eqnarray}
where $\|\cdot\|$ is the $L_2$ norm.
In their seminal paper, Ke and Wang consider a penalized maximum likelihood approach for the estimation of $(\bdtheta,f)$. That is, they propose to solve
\begin{equation}\label{pen}
\max_{\bdtheta,f } \left\{\ell(\bdy; (\bdtheta,f)) - n\lambda J(f)\right\}
\end{equation}
where $\ell(\bdy; (\bdtheta,f))$ is the marginal log-likelihood, $J(f)$ is some roughness penalty and $\lambda$ is a smoothing parameter. Moreover, they assume that $f$ belongs to some reproducing kernel Hilbert space (RKHS) ${\cal H}= {\cal H}_1 \oplus {\cal H}_2$, where ${\cal H}_1$ is a finite dimensional space of functions, ${\cal H}_1= span \{\psi_1,\dots,\psi_M\}$, and ${\cal H}_2$ is a RKHS itself (see Section~2 of \cite{KeWang}). Since the nonlinear function $f$ interacts in a complicated way  with the random effects and the integral in (\ref{lik}) is intractable, they replace $\ell(\bdy; (\bdtheta,f))$ by a linear Laplace approximation $\tilde{\ell}(\bdy; (\bdtheta,f,\tilde{\bdphi}))$, where $\tilde{\bdphi}$ is some convenient value for $\bdphi$ (see (10) in Section~3.1 of \cite{KeWang}). Then, they propose to estimate $(\bdtheta, f)$
with the following iterative procedure:
\begin{enumerate}
\item[i)] given an estimate of $f$, get estimates of $\bdtheta$ and $\bdphi$ by fitting the resultant nonlinear mixed model by linearizing the log-likelihood (replacing $\ell$ by $\tilde{\ell}$). Indeed, in practice they use the S-PLUS function {\tt{nlme}}, \cite{Pin00}, to solve this step.
\item[ii)] given an estimate of $\bdtheta$, $\hat{\bdtheta}$, estimate $f$ as the solution to
\end{enumerate}
$$\hspace{-0.5cm}\max_{f\in{\cal H}} \left\{\ell(\bdy; (\hat{\bdtheta},f)) - n\lambda J(f)\right\} \approx \max_{f\in{\cal H}}\left\{\tilde{\ell}(\bdy; (\hat{\bdtheta},f,\tilde{\bdphi})) - n\lambda J(f)\right\} = \max_{f\in \tilde{{\cal W}}_1}\left\{\tilde{\ell}(\bdy; (\hat{\bdtheta},f,\tilde{\bdphi})) - n\lambda J(f)\right\},$$

where $\tilde{{\cal W}}_1$ is some finite dimensional space whose particular definition depends on the set of points $\{c(\tilde{\bdphi}_i; \bdx_{ij}), i=1,\dots,N,\,\, j=1,\dots,n_i\}$ in which the function $f$ is evaluated. Indeed, since the approximated log-likelihood involves a bounded linear functional, the maximizer in ${\cal H}$ of $\tilde{\ell}(\bdy; (\hat{\bdtheta},f,\tilde{\bdphi})) - N\lambda J(f)$ belongs to $\tilde{{\cal W}}_1$ (see Section~ 4.1 of \cite{KeWang} and \cite{Wang98}). However, as it is pointed out by Lin and Zhang in their comment to \cite{KeWang}, the solution to the original problem, namely (\ref{pen}), in such a space ${\cal H}$ might not exist, and if it exists, it may lie in an infinite dimensional space and might not be unique. This is the main difference with standard regression models in which the maximizer in ${\cal H}$ of the penalized log-likelihood belongs to a finite dimensional space (see \cite{Wahba} for instance). This result also holds for particular nonlinear nonparametric regression models (see \cite{KeWang04}), but cannot be generally extended to SNMMs because of the interaction between the random effects and the nonlinear function $f$.\\
So in fact, the approach of Ke and Wang consists in choosing $\tilde{{\cal W}}_1$ as a finite-dimensional approximation of ${\cal H}$ to solve (\ref{pen}).\\
Also, it is important to point out some drawbacks of the approximated methods based on linearization of the log-likelihood, such as the Laplace's approximation used by Ke and Wang. It has been shown that they can produce inconsistent estimates of the fixed effects, in particular when the number of measurements per subject is not large enough (\cite{RamosPantula,Vonesh96}). In addition, simulation studies have shown unexpected increases in the type I error of the likelihood ratio and Wald tests based on these linearization methods (\cite{DingWu}).\\

In this paper we propose an alternative estimation procedure in SNMMs. On the one hand, for the parametric step we will focus on the maximization of the exact likelihood. We propose to use a stochastic version of the EM algorithm, the so-called SAEM algorithm introduced by \cite{saem} and extended by \cite{Kuhn05} for nonlinear mixed models, to estimate $\bdtheta$ without any approximation or linearization. This stochastic EM algorithm replaces the usual E step of EM algorithm \citep{EM} by a simulation step and a stochastic procedure, and converges to a local maximum of the likelihood. The SAEM has been proved to be computationally much more efficient than other stochastic algorithms as for example the classical Monte Carlo EM (MCEM) algorithm \citep{Wei90} thanks to a recycling of the simulated variables from one iteration to the next (see \cite{Kuhn05}). Indeed, previous attempts to perform exact ML estimation in SNMMs have been discarded because of the computational problems related to the use of an MCEM algorithm (see \cite{LiuWu07,LiuWu08,LiuWu09}).  Moreover we use a Restricted Maximum Likelihood (REML) version of the SAEM algorithm to correct bias estimation problems of the variance parameters following the same strategy as \cite{Cristian1}.\\
On the other hand, for the nonparametric step we will propose a LASSO-type method for the estimation of $f$. The popular LASSO estimator (least absolute shrinkage and selection operator, \cite{LASSO}) based on $\ell_1$ penalized least squares, has been extended in the last years to nonparametric regression (see for instance \cite{BRT}). It has been also used by \cite{vdgeer} in  high-dimensional linear mixed-effects models.
In the nonparametric context, the idea is to reconstruct a sparse approximation of $f$ with linear combinations of elements of a given set of functions $\{f_1,\dots, f_M\}$, called dictionary. That is, we are implicitly assuming that $f$ can be well approximated with a small number of those functions. In practice, for the nonparametric regression problem, the dictionary can be a collection of basis functions from different bases (splines with fixed knots, wavelets, Fourier, etc.). The advantage of this approach with respect to the penalized maximum likelihood estimation in an approximate space of functions, as proposed by Ke and Wang (2001), is that now the selection of the finite-dimensional space among a large collection of possible spaces spanned by very different functions is automatic and based on data.  
This approach allows to construct a good approximation of the nonparametric function which is sparse thanks to the large dictionary. The sparsity of the approximation gives a model more interpretable and since few coefficients have to be estimated, this minimizes the estimation error.
The LASSO algorithm allows to use the dictionary approach to select a sparse approximation, unlike to  wavelet thresholding or  $\ell_0$- penalization. Moreover the LASSO algorithm has a low computational cost since it is based on a convex penalty. 

We can summarize our iterative estimation procedure as:
\begin{enumerate}
\item[i)] given $\hat{f}$, an estimate of $f$, get estimates of $\bdtheta$ and $\bdphi$ by fitting the resulting nonlinear mixed model with the SAEM algorithm (using ML or REML method).
\item[ii)] given estimates of $\bdtheta$ and $\bdphi$, solve the resulting nonparametric regression problem using a LASSO-type method.
\end{enumerate}
\medskip

In fact, since the SAEM algorithm is an iterative procedure itself, instead of running the whole SAEM algorithm until convergence for each given $f$ at step i), we will rather perform only one iteration of the algorithm in order to update the $\bdtheta$ and $\bdphi$ estimates from the current value of $\hat{f}$. Then, the nonparametric estimation of step ii) will be performed at each iteration of the SAEM algorithm, as we will see in Section \ref{algos}.\\

The rest of the article is organized as follows. In Section \ref{par_est} we describe the SAEM algorithm and its REML version in the framework of SNMMs. In Section \ref{f_est} we propose a LASSO-type method for the estimation of $f$ in the resulting nonparametric regression problem after estimation of $\bdtheta$ and $\bdphi$. We derive oracle inequalities and subset selection properties for the proposed estimator. In Section \ref{algos}, we describe the algorithm that combines both procedures to perform joint estimation of $(\bdtheta, f)$ in the SNMM. Finally, in Section \ref{applications}, we illustrate our method through simulated and real data. We conclude the article in Section \ref{con}. The proofs of the results of Section \ref{f_est} are in the Appendix.

\section{Estimation of the finite-dimensional parameters}
\subsection{SAEM estimation of $\bdtheta$ and $\bdphi$}\label{par_est}

Let us focus on the first point of our procedure, which is performed by the Stochastic Approximation EM algorithm, SAEM (\cite{saem}). In this subsection we consider that we have an estimate of $f$, $\hat{f}$, obtained in the previous estimation step that does not change during the estimation of $\bdtheta$. Thus, we can proceed as if $f$ was a known nonlinear function and we fall into the SAEM estimation of nonlinear mixed-effects model framework (see \cite{Kuhn05}). In fact, note that since the estimation of $f$ is performed by solving a nonparametric regression problem with regression variables  $c(\hat{\bdphi}_i; \bdx_{ij}), i=1,\dots,N,\,\, j=1,\dots,n_i$ (see Section~\ref{f_est}), it will depend on the estimated value of $\bdphi$ at the precedent iteration. Then, we will note $\hat{f}_{-}$ the current estimated function.\\
The complete likelihood for model (\ref{modmixte}) is:
\begin{eqnarray*}\label{compl}
p(\bdy,\bdphi; \bdtheta)&=& p(\bdy|\bdphi; \bdtheta) p(\bdphi; \bdtheta)\\
&=&\dfrac{1}{(2 \pi \sigma^2)^{\frac{n}{2}}} \exp\left\{\dfrac{-1}{2\sigma^2} \|\bdy-\bdg(\bdphi,\hat{f}_{-})\|^2\right\}
\dfrac{1}{(2 \pi)^{\frac{Np}{2}} |\bdGamma|^{\frac{N}{2}}} \exp\left\{\dfrac{-1}{2}\|\widetilde{\bdGamma}^{-1/2}(\bdphi-\bdA \bdbeta)\|^2\right\}\nonumber\\
&=&\frac{1}{(2 \pi)^{\frac{n+Np}{2}} (\sigma^2)^{\frac{n}{2}} |\bdGamma|^{\frac{N}{2}}}\exp\left\{\dfrac{-1}{2} \left(\frac{1}{\sigma^2} \|\bdy-\bdg(\bdphi,\hat{f}_{-})\|^2 + \|\widetilde{\bdGamma}^{-1/2}(\bdphi-\bdA \bdbeta)\|^2\right)\right\}
\end{eqnarray*}
where $n=\sum_{i=1}^N n_i$.
The complete log-likelihood is:
\begin{equation}\label{compl_log}
\log p(\bdy,\bdphi; \bdtheta)= \dfrac{-1}{2}\left\{C+n\log\sigma^2 +N \log  |\bdGamma| + \frac{1}{\sigma^2} \|\bdy-\bdg(\bdphi,\hat{f}_{-})\|^2 + \|\widetilde{\bdGamma}^{-1/2}(\bdphi-\bdA \bdbeta)\|^2\right\}
\end{equation}
where $C$ is a constant that does not depend on $\bdtheta$.

The principle of the EM algorithm, \cite{EM}, is to maximize at iteration $k$ the conditional expectation of $\log p(\bdy,\bdphi; \bdtheta)$ given the observed data and the precedent value of $\bdtheta$, $\bdtheta^{(k)}$, that is
\begin{equation*}\label{Q}
Q_{k+1}(\bdtheta)= \E\left( \log p(\bdy,\bdphi; \bdtheta) | \bdy; \bdtheta^{(k)} \right).
\end{equation*}
This can be simplified if we assume that the distribution of the complete-data model belongs to the exponential family, that is, if
\begin{equation*}\label{S}
\log p(\bdy,\bdphi; \bdtheta) = - \Psi(\bdtheta) + \langle S(\bdy,\bdphi),\Phi(\bdtheta)\rangle \end{equation*}
where $\langle \cdot,\cdot\rangle$ stands for the scalar product and $S(\bdy,\bdphi)$ is the sufficient statistics of the complete-data model. In that case, the EM algorithm consists in iterating the two following steps:
\begin{itemize}
\item[-] E step: evaluate the quantity $s_{k+1}= \E [S(\bdy,\bdphi)| \bdy; \bdtheta^{(k)}]$.
\item[-] M step: update the value of $\bdtheta$: $\bdtheta^{(k+1)}= \mbox{arg} \max_{\bdtheta} \{- \Psi(\bdtheta)+ \langle s_{k+1},\Phi(\bdtheta)\rangle \}$.
\end{itemize}
One of the main drawbacks of the EM algorithm is that the computation in the E step is intractable in many cases. The SAEM algorithm replaces, at each iteration, the step E by a simulation step (S) of the missing data ($\bdphi$) and an approximation step (A) of $Q_{k+1}(\bdtheta)$. Then, at iteration $k$, the SAEM algorithm can be written as:\vspace{0.2cm}\\
\fbox{{\parbox{1\linewidth}{
\begin{itemize}
\item[-] S step: simulate $m$ values of the random effects, $\bdphi^{(k+1,1)},\ldots,\bdphi^{(k+1,m)}$, from the conditional law $p(\cdot |\bdy; \bdtheta^{(k)}) $.
\item[-] A step: update $s_{k+1}$ according to: $s_{k+1}=s_k + \gamma_k \left[ \dfrac{1}{m}\displaystyle\sum_{l=1}^{m}S(\bdy,\bdphi^{(k+1,l)}) - s_k\right] $.
\item[-] M step: update the value of $\bdtheta$: $\bdtheta^{(k+1)}= \mbox{arg} \max_{\bdtheta} \{- \Psi(\bdtheta)+ \langle s_{k+1},\Phi(\bdtheta)\rangle \}\vspace{-0.9cm}$. \begin{equation}\label{saem-ml}\end{equation}
\end{itemize}
}}}
\vspace{0.2cm}\\
The sequence $\{s_k\}$ is initialized at $s_0$ and $\gamma_k$ is a decreasing sequence of positive numbers, as presented by \cite{Kuhn04}, which accelerates the convergence.

For the approximation and the maximization steps, we need to define the quantities $s_k$. From (\ref{compl_log}), we have that
\begin{equation*}\label{compl_log_suite}
\log p(\bdy,\bdphi; \bdtheta)= -\dfrac{1}{2}\left\{C+n\log\sigma^2 +N  \log  |\bdGamma| + \frac{1}{\sigma^2}
\|\bdy-\bdg(\bdphi,\hat{f}_{-})\|^2 + \sum_{i=1}^N (\bdphi_i-\bdA_i \bdbeta)' {\bdGamma}^{-1}(\bdphi_i-\bdA_i \bdbeta)\right\}.
\end{equation*}
Then, the aproximation step reduces to updating the sufficient statistics for the complete model
\begin{eqnarray*}
s_{1,i,k+1}&=& s_{1,i,k} + \gamma_k \left[ \dfrac{1}{m}\displaystyle\sum_{l=1}^{m}\bdphi_i^{(k+1,l)} - s_{1,i,k}\right],\quad i=1,\dots,N \nonumber \\
s_{2,k+1}&=&  s_{2,k} + \gamma_k \left[ \dfrac{1}{m}\displaystyle\sum_{l=1}^{m}\sum_{i=1}^{N}\bdphi_i^{(k+1,l)}\bdphi_i^{(k+1,l)'} - s_{2,k}\right]\nonumber \\
s_{3,k+1}&=& s_{3,k}+ \gamma_k \left[ \dfrac{1}{m}\displaystyle\sum_{l=1}^{m}\|\bdy-\bdg(\bdphi^{(k+1,l)},\hat{f}_{-})\|^2 - s_{3,k}\right].
\end{eqnarray*}
Now, $\bdtheta^{(k+1)}$ is obtained in the maximization step as follows:
\begin{eqnarray*}
\bdbeta^{(k+1)}&=&  \left(\sum_{i=1}^{N} \bdA_i'{\bdGamma}^{(k)^{-1}}\bdA_i \right)^{-1} \sum_{i=1}^{N} \bdA_i'{\bdGamma}^{(k)^{-1}} s_{1,i,k+1}  \nonumber \\
\bdGamma^{(k+1)}&=& \dfrac{1}{N}\! \left(\! s_{2,k+1} -\!\sum_{i=1}^{N} \bdA_i \bdbeta^{(k+1)} s_{1,i,k+1}' -\!\sum_{i=1}^{N} s_{1,i,k+1}\left(\!\bdA_i \bdbeta^{(k+1)}\right)' + \!\sum_{i=1}^{N} \bdA_i \bdbeta^{(k+1)}\left(\!\bdA_i \bdbeta^{(k+1)}\right)' \! \right)\nonumber \\
\sigma^{2^{(k+1)}}&=&\dfrac{s_{3,k+1}}{n}.
\end{eqnarray*}

When the simulation step cannot be directly performed, \cite{Kuhn04} propose to combine this algorithm with a Markov Chain Monte Carlo (MCMC) procedure. Then, the simulation step becomes:
\begin{itemize}
\item[-] S step: using $\bdphi^{(k,l)}$, draw $\bdphi^{(k+1,l)}$ with transition probability $\Pi_{\bdtheta^{(k)}}(\cdot \vert \bdphi^{(k,l)})$, $l=1,\ldots,m$,
\end{itemize}
that is, $(\bdphi^{(k+1,1)}),\ldots,(\bdphi^{(k+1,m)})$ are $m$ Markov chains with transition kernels $\left(\Pi_{\bdtheta^{(k)}}\right)$. In practice, these Markov chains are generated using a Hastings-Metropolis algorithm (see \cite{Kuhn05} for details).\\
With respect to the number of chains, the convergence of the whole algorithm to a local maximum of the likelihood is granted even for $m=1$. Greater values of $m$ can accelerate the convergence, but in practice $m$ is always lower than 10. This is the main difference with the MCEM algorithm, in which very large samples of the random effects have to be generated in order for the algorithm to converge.

\subsection{REML estimation of variance components}\label{REML}
It is well known that the maximum likelihood estimator of variance components in mixed effects models can be biased downwards because it does not adjust for the loss of degrees of freedom caused by the estimation of the fixed effects. This is also true in the context of SNMMs as \cite{KeWang} point out in their paper.\\
Restricted maximum likelihood (REML), as originally formulated by \cite{REML1} in the context of linear models, is a method that corrects this problem by maximizing the likelihood of a set of linear functions of the observed data that contain none of the fixed effects of the model. But this formulation does not directly extend beyond linear models, where in general it is not possible to construct linear functions of the observed data that do not contain any of the fixed effects. However, in the case of nonlinear models, other alternative formulations of REML have been proposed. Here, we will consider the approach of \cite{REML2}, that consists in the maximization of the likelihood after integrating out the fixed effects. The combination of this REML approach with the SAEM algorithm in the context of nonlinear mixed effects models has been studied recently by \cite{Cristian1}. The authors showed the efficiency of the method against purely ML estimation performed by SAEM and against REML estimation based on likelihood approximation methods.

Then, following the ideas of \cite{Cristian1}, we will note $\bdz=(\bdphi, \bdbeta)$ the random effects and $\tilde{\bdtheta}=(\bdGamma,\sigma^2)$ the new parameter of the model. As in the general case, the simulation step is performed through an MCMC procedure. Here, since we have to draw values from the joint distribution of $(\bdphi, \bdbeta)\vert \bdy;\tilde{\bdtheta}^{(k)}$, we use a Gibbs scheme, i.e., we iteratively draw values from the conditional distributions of $\bdphi\vert \bdy, \bdbeta^{(k)};\tilde{\bdtheta}^{(k)}$ and $\bdbeta\vert \bdy,\bdphi^{(k)};\tilde{\bdtheta}^{(k)}$. Then, we use again a Hastings-Metropolis algorithm to obtain approximations of these conditional distributions.\\
Finally, iteration $k$ of the SAEM-REML algorithm for model (\ref{mod_vec}) writes:\vspace{0.2cm}\\
\fbox{{\parbox{1\linewidth}{
\begin{itemize}
\item[-] S step: using $\bdz^{(k,l)}=(\bdphi^{(k,l)}, \bdbeta^{(k,l)})$, simulate $\bdz^{(k+1,l)}=(\bdphi^{(k+1,l)}, \bdbeta^{(k+1,l)})$, $l=1,\ldots,m$ with a Metropolis-within-Gibbs scheme.
\item[-] A step: update $\tilde{s}_{k+1}$ according to $\,\,\,\tilde{s}_{k+1}\!\!=\tilde{s}_k + \gamma_k \left[ \dfrac{1}{m}\displaystyle\sum_{l=1}^{m}\tilde{S}(\bdy,\bdz^{(k+1,j)}) - \tilde{s}_k\right] $, namely:
\begin{eqnarray}\label{saem-reml}
\tilde{s}_{1,k+1}&=&  \tilde{s}_{1,k} + \gamma_k \left[ \dfrac{1}{m}\displaystyle\sum_{l=1}^{m}\sum_{i=1}^{N}\bdeta_i^{(k+1,l)}\bdeta_i^{(k+1,l)'} - \tilde{s}_{1,k}\right]\nonumber \\
\tilde{s}_{2,k+1}&=& \tilde{s}_{2,k}+ \gamma_k \left[ \dfrac{1}{m}\displaystyle\sum_{l=1}^{m}\|\bdy-\bdg(\bdz^{(k+1,l)},\hat{f}_{-})\|^2 - \tilde{s}_{2,k}\right]
\end{eqnarray}
where $\bdeta_i^{(k+1,l)}=\bdphi_i^{(k+1,l)} - A_i\bdbeta^{(k+1,l)}$.
\item[-] M step: update the value of $\tilde{\bdtheta}$ by $\,\,\,\tilde{\bdtheta}^{(k+1)}\!\!= \mbox{arg} \max_{\tilde{\bdtheta}} \{- \Psi(\tilde{\bdtheta})+ \langle \tilde{s}_{k+1},\Phi(\tilde{\bdtheta})\rangle \}$, namely:
$$\bdGamma^{(k+1)}= \dfrac{\tilde{s}_{1,k+1}}{N}  \quad \quad \mbox{and} \quad \quad \sigma^{2^{(k+1)}}=\dfrac{\tilde{s}_{2,k+1}}{n}.$$
\end{itemize}
}}}

\section{Estimation of the function $f$ using a LASSO-type method}\label{f_est}
\subsection{Estimation procedure}\label{f_alg}
In this part, our objective is to estimate $f$ in the model (\ref{modmixte}) using the observations $y_{i,j}$ and assuming that for $i=1,\ldots,N$ we have $\bdphi_i=\hat{\bdphi}_i$ and $\sigma^2=\hat{\sigma}^2$ where the estimates $\hat{\bdphi}_i$ and $\hat{\sigma}^2$ have been obtained in the precedent SAEM step. Since $g$ satisfies (\ref{formef}), model (\ref{modmixte}) can be rewritten as
$$\tilde{y}_{ij}=b(\bdphi_i; \bdx_{ij})f(\tilde{\bdx}_{ij})+\varepsilon_{ij},\quad i=1\dots,N, \,\, j=1,\dots,n_i $$
with $\tilde{y}_{ij}=y_{ij}-a(\bdphi_i; \bdx_{ij})$ and $\tilde{\bdx}_{ij}=c(\bdphi_i; \bdx_{ij})$. Of course, since the $\hat\bdphi_i$'s and $\hat\sigma^2$ depend on the observations, the distribution of  $\hat\sigma^{-1}\tilde{y}_{ij}$ is no longer Gaussian. But in the sequel, to be able to derive theoretical results, we still assume that
\begin{equation}\label{norm}
\varepsilon_{ij}\stackrel{iid}{\sim} {\cal N}(0, \sigma^2),
\end{equation}
where the value of $\sigma^2$ is given by $\hat\sigma^2$. Simulation studies of Section~\ref{applications} show that this assumption is reasonable. However, note that (\ref{norm}) is true at the price of splitting the data set into two parts: the first part for estimating $\bdtheta$ and $\bdphi$, the second part for estimating $f$.
Now, reordering the observations, it is equivalent to observing $(y_1,\ldots,y_n)$ with $n=\sum_{i=1}^N n_i$, such that
\begin{equation}\label{modregLASSO}
y_i=b_if(x_i)+\e_i,\quad\varepsilon_{i}\sim {\cal N}(0, \sigma^2) \mbox{ i.i.d.}
\end{equation}
where the $b_i$'s and the design $(x_i)_{i=1,\ldots,n}$ are known and depend on the estimators of the precedent SAEM step and the $\e_i$'s are 
random variables with variance $\sigma^2$ estimated by $\hat\sigma^2$. Note that the notation $y_i$, $i=1,\ldots,n$, does not correspond to the original observations in the SNMM or to any of the values introduced in the previous sections, and it is used in this section for the sake of simplicity. Without loss of generality, we suppose that $b_i\neq 0$ for all $i=1,\ldots,n$.

In the sequel, our objective is then to estimate $f$ nonparametrically in model (\ref{modregLASSO}). A classical method would consist in decomposing  $f$ on an orthornormal basis (Fourier basis, wavelets,...) and then to use a standard nonparametric procedure to estimate the coefficients of $f$ associated with this basis ($\ell_0$-penalization, wavelet thresholding,...). In the same spirit as \cite{berlepenrivo} who investigated the problem of density estimation, we wish to combine a more general  dictionary approach with an estimation procedure leading to fast algorithms. The dictionary approach consists in proposing estimates that are linear combinations of various types of functions. Typically, the dictionary is built by  gathering together atoms of various classical orthonormal bases. This approach offers two advantages. First, with a more wealthy dictionary than a classical orthonormal basis, we aim at obtaining sparse estimates leading to few estimation errors of the coefficients. Secondly, if the estimator is sparse enough, interesting interpretations of the results are possible by using the set of the non-zero coefficients, which corresponds to the set of functions of the dictionary "selected" by the procedure. For instance, we can point out the frequency of periodic components of the signal if trigonometric functions are selected or local peaks if some wavelets are chosen by the algorithm. Both aspects are illustrated in the next sections. $\ell_0$-penalization or thresholding cannot be combined with a dictionary approach if we wish to obtain fast and good algorithms. But LASSO-type estimators based on $\ell_1$-penalization, leading to minimization of convex criteria, constitute a natural tool for the dictionary approach. Furthermore, unlike ridge penalization or more generally $\ell_p$-penalization with $p>1$, $\ell_1$-penalization leads to sparse solutions for the minimization problem, in the sense that if the tuning parameter is large enough some coefficients are exactly equal to 0 (see \cite{LASSO}).

There is now a very huge literature on LASSO-type procedures. From the theoretical point of view and in the specific context of the regression model close to (\ref{modregLASSO}), we mention that LASSO procedures have already been studied by  \cite{btw06}, \cite{btw}, \cite{Gau},   \cite{bunea2}, \cite{BRT}, \cite{HighSara}, and  \cite{livreBulmannSara} among others.

In our setting, the proposed procedure is the following. For $M\in\Ne^*$, we consider a set of functions $\{\p_1,\ldots,\p_M\}$, called the {\it dictionary}. We denote for $\lambda\in\R^M$, $$f_\lambda=\sum_{j=1}^M \lambda_j\p_j.$$
Our objective is to find good candidates for estimating $f$ which are linear combinations of functions of the dictionary, i.e. of the form $f_\lambda$.
We consider, for $\lambda\in \R^M$
\begin{equation*}
\mbox{crit} (\lambda)= \frac{1}{n} \sum_{i=1}^n \left(y_i-b_i f_\lambda(x_i)\right)^2+2\sum_{j=1}^M r_{n,j}|\lambda_j|,
\end{equation*}
where
$r_{n,j}=\sigma\|\p_j\|_n\sqrt{\frac{\gamma \log M}{n}}$ with $\gamma>0$ and for a function $h$
$$\|h\|_n^2=\frac{1}{n}\sum_{i=1}^nb_i^2h^2(x_i).$$
We call the LASSO estimator $\hat{\lambda}$ the minimizer of $\la\longmapsto\mbox{crit}(\lambda)$ for $\lambda\in\R^M$ and we denote $\hat{f}=f_{\hat{\lambda}}$.

The function $\lambda\longmapsto\mbox{crit} (\lambda)$ is the sum of two terms: the first one is a goodness-of-fit criterion based on the $\ell_2$-loss and the second one is a penalty term that can be viewed as the weighted $\ell_1$-norm of $\lambda$.

Before going further, let us discuss the important issue of  tuning. In our context, the tuning parameter is the constant $\gamma$. From a theoretical point of view (see Theorem~\ref{oracleadmi}), the benchmark value for $\gamma$ is $2$. In the sequel, $\gamma$ will be chosen satisfying two criteria: to be as close as possible to this benchmark value and allowing the stability of the SAEM algorithm. In Section~\ref{applications}, we will see that sometimes we choose values of $\gamma$ smaller than $2$ but relatively close of it, in particular to obtain the convergence of the variance components estimates, which is always challenging in NLME models.

Once we have chosen a value for $\gamma$ satisfying these two criteria, the numerical scheme of the nonparametric step is the following:\vspace{0.2cm}\\
\fbox{{\parbox{1\linewidth}{
\begin{itemize}
\item[-] Using the estimates of the $\phi_i$'s and of $\sigma^2$ obtained in the previous iteration of SAEM, compute for $i=1,\ldots,n$, the observations $y_i$, the constants $b_i$ and the design $x_i$.
\item[-] Evaluate the dictionary $\{\p_1,\ldots,\p_M\}$ at the design and calculate $r_{n,j}$.
\item[-] Obtain the LASSO estimates $\hat{\lambda}$ and $f_{\hat{\lambda}}$.
\end{itemize}
}}}
\vspace{0.2cm}\\
In practice, there exist many efficient algorithms to tackle this third point, namely, the minimization on $\lambda$ of $crit(\lambda)$. For the implementation of our estimation procedure we have considered the approach used by \cite{berlepenrivo} which consists in using the LARS algorithm.
\subsection{Theoretical results}
Numerical results of our procedure are presented in next sections but we now validate our approach from a theoretical point of view. More precisely, we consider the oracle approach.
\subsubsection{Assumptions}
As usual, assumptions on the dictionary are necessary to obtain oracle results for LASSO-type procedures. We refer the reader to \cite{LassoAss} for a good review of different assumptions considered in the literature for LASSO-type estimators and connections between them. The dictionary approach aims at extending results for orthonormal bases. Actually, our assumptions express the relaxation of the orthonormality property. To describe them, we introduce the following notation.
For $l\in\Ne$, we denote
\begin{align*}
\nu_{\min}(l)&
=\min_{|J|\leq l}\min_{\substack{\lambda\in\R^M\\\lambda_J\neq0}}
\frac{\normD{f_{\lambda_J}}^2}{\normd{\lambda_J}^2}&\text{and}&&
\nu_{\max}(l)&
=\max_{|J|\leq l}\max_{\substack{\lambda\in\R^M\\\lambda_J\neq0}}
\frac{\normD{f_{\lambda_J}}^2}{\normd{\lambda_J}^2},
\end{align*}
where $\normd{\cdot}$ is the $l_2$ norm in $\R^M$.
The notation $\lambda_J$ means that for any $k\in\{1,\ldots,M\}$,  $(\lambda_J)_k=\lambda_k$ if $k\in J$ and $(\lambda_J)_k=0$ otherwise. Previous quantities correspond to the ``restricted'' eigenvalues of the Gram matrix $G=(G_{j,j'})$ with coefficients
$$G_{j,j'}=\frac{1}{n}\sum_{i=1}^n b_i^2\p_j(x_i)\p_{j'}(x_i).$$
Assuming that $\nu_{\min}(l)$ and $\nu_{\max}(l)$ are close to 1 means that every set of columns of $G$ with cardinality less than $l$  behaves like an orthonormal system. We also consider the restricted correlations
\[
\delta_{l,l'}=
\max_{\substack{\ |J|\le l\\\ |J'|\le l'\\J\cap
    J'=\emptyset}}
\max_{\substack{\lambda,\lambda'\in \R^M\\
\lambda_J\neq0,\lambda'_{J'}\neq0
}
}
 \frac{\langle f_{\lambda_{J}}, f_{\lambda'_{J'}} \rangle}
{\normd{\lambda_{J}}\normd{\lambda'_{J'}}},
\]
where $\langle f, g \rangle=\frac{1}{n}\sum_{i=1}^nb_i^2f(x_i)g(x_i)$.
Small values of $\delta_{l,l'}$ means that two disjoint sets of columns of $G$ with cardinality less than $l$ and $l'$ span nearly orthogonal spaces. We will use the following assumption considered in \cite{BRT}.
\begin{assump}
For some integer $1\le s\le M/2$, we have
\begin{equation}\label{Ass1}
\nu_{\min}(2s)> \delta_{s,2s}.\tag{\mbox{A1}(s)}
\end{equation}
\end{assump}
Oracle inequalities of the Dantzig selector were established under
this assumption in the parametric linear model by \cite{candes} and for density estimation by \cite{berlepenrivo}. It was also considered by \cite{BRT} for nonparametric regression and for the LASSO estimate.\\

\noindent Let us denote
\[
\kappa_s =\sqrt{\nu_{\min}(2s)}\left(1-\frac{\delta_{s,2s}}{\nu_{\min}(2s)}\right)>0,\quad
\mu_s=\frac{\delta_{s,2s}}{\sqrt{\nu_{\min}(2s)}}.
\]
We will say that $\lambda\in\R^M$ satisfies the Dantzig constraints if for all $j=1,\ldots,M$
\begin{equation}\label{dantzig}
\left|(G\lambda)_j-\hat{\beta_j}\right|\le r_{n,j},
\end{equation}
where $$\hat{\beta}_j=\frac{1}{n}\sum_{i=1}^nb_i \p_j(x_i)Y_i.$$
We denote $\mathcal{D}$ the set of $\lambda$ that satisfies (\ref{dantzig}). The classical use of Karush-Kuhn-Tucker conditions shows that the LASSO estimator $\hat{\lambda}\in{\mathcal D}$, so it satisfies the Dantzig constraint.
\subsubsection{Oracle inequalities}
We obtain the following oracle inequalities.
\begin{Th}\label{oracleadmi}
Let $\gamma>2$. With probability
at least $1-M^{1-\gamma/2}$,
for any  integer $s<n/2$  such
that (\ref{Ass1}) holds,
 we have for any $\betamod>0$,
\begin{equation}\label{oracle1}\normD{\hat{f}-\fo}^2 \leq \inf_{\lambda\in\R^M}
\inf_{\substack{J_0 \subset \{1,\ldots,M\}\\
  |J_0|=s}} \left\{ \normD{f_\lambda-\fo}^2+\betamod \left(1+\frac{2\mu_s}{\kappa_s}\right)^2\frac{
\Lam^2}{s}+16s
\left(\frac{1}{\betamod}+\frac{1}{\kappa_s^2}\right)
r_n^2 \right\}
\end{equation}
where
$$r_n=\sup_{j=1,\ldots,M}r_{n,j},$$
\[
\Lam = \normu{\lambda_{\Jo^C}}+\frac{\left(\normu{\hat
  \lambda}-\normu{\lambda}\right)_+}{2},
\]
for any $x\in\R$ $x_+:=\max(x,0)$ and $\normu{\cdot}$ is the $l_1$ norm in $\R^M$.
\end{Th}

\begin{Th}\label{oracleadmi2}
Let $\gamma>2$. With probability
at least $1-M^{1-\gamma/2}$,
for any  integer $s<n/2$  such
that (\ref{Ass1}) holds,
 we have for any $\betamod>0$,
\begin{equation}\label{oracle2}\normD{\hat{f}-\fo}^2 \leq \inf_{\lambda\in\mathcal{D}}
\inf_{\substack{J_0 \subset \{1,\ldots,M\}\\
  |J_0|=s}} \left\{ \normD{f_\lambda-\fo}^2+\betamod \left(1+\frac{2\mu_s}{\kappa_s}\right)^2\frac{\normu{\lambda_{\Jo^C}}+\normu{\hat{\lambda}_{\Jo^C}}
}{s}+32s
\left(\frac{1}{\betamod}+\frac{1}{\kappa_s^2}\right)
r_n^2 \right\}.
\end{equation}
\end{Th}

Similar oracle inequalities were established by \cite{btw06}, \cite{btw}, \cite{Gau}, or \cite{HighSara}. But in these works, the
functions of the dictionary are assumed to be bounded by a constant independent of $M$ and $n$.
Let us comment the right-hand side of inequalities (\ref{oracle1}) and (\ref{oracle2}) of Theorems~\ref{oracleadmi} and \ref{oracleadmi2}. The first term is an approximation term which measures the closeness between $f$ and $f_\lambda$ and that can vanish if $f$ is a linear combination of the functions of the dictionary. The second term can be considered as a bias term. In both theorems, the term $\normu{\lambda_{\Jo^C}}$  corresponds to the cost of having $\lambda$ with a support different of $J_0$. For a given $\lambda$, this term can be minimized by choosing $J_0$ as the set of largest coordinates of $\lambda$. Note that if the function $f$ has a sparse expansion on the dictionary, that is $f=f_\lambda$ where $\lambda$ is a vector with $s$ non-zero coordinates, then by choosing $J_0$ as the set of the $s$ non-zero coordinates, the approximation term and the term $\normu{\lambda_{\Jo^C}}$ vanish. In Theorem~\ref{oracleadmi}, the term $\left(\normu{\hat
  \lambda}-\normu{\lambda}\right)_+$ will be smaller as the $\ell_1$-norm of the LASSO estimator is small and this term is equal to 0 if $\normu{\hat  \lambda}\leq\normu{  \lambda}$, which is frequently the case. In Theorem~\ref{oracleadmi2}, given a vector $\lambda$ such that $f_\lambda$ approximates well $f$, the term $\normu{\hat{\lambda}_{\Jo^C}}$ will be small if the LASSO estimator selects the largest coordinates of $\lambda$. The last term can be viewed as a variance term corresponding to the estimation of $f$ as linear combination of $s$ functions of the dictionary (see \cite{berlepenrivo} for more details). Finally, the parameter $\alpha$ calibrates the weights given for the bias and variance terms.

The following section deals with estimation of sparse functions.

\subsubsection{The support property of the LASSO estimate}\label{supp-sect}
Let $\gamma>2$. In this section, we apply the LASSO procedure with $\tilde r_{n,j}$ instead of $r_{n,j}$, with
\[\tilde r_{n,j}=\sigma\|\p_j\|_n\sqrt{\frac{\tilde\gamma \log M}{n}},\quad\tilde\gamma>\gamma.\]
We assume that the regression function $\fo$ can be decomposed on the dictionary: there exists $\lambda^*\in\R^M$ such that
\[\fo=\sum_{j=1}^M\la^*_j\p_j.\]
We denote $S^*$ the support of $\lambda^*$:
\[S^*=\left\{j\in\{1,\ldots,M\}:\quad\lambda^*_j\not=0\right\},\]
and by $s^*$ the cardinal of $S^*$.
We still consider the LASSO estimate $\hat\lambda$ and, similarly, we denote $\hat S$ the support of $\hat\la$:
\[\hat S=\left\{j\in\{1,\ldots,M\}:\quad\hat\lambda_j\not=0\right\}.\]
One goal of this section is to show that with high probability, we have:
\[\hat S\subset S^*.\]
We have the following result.
\begin{Th}\label{theosupport}
We define
\[\rho(S^*)=\max_{k\in S^*}\max_{ j\not=k}\frac{|<\p_j,\p_k>|}{\|\p_j\|_n\|\p_k\|_n}\]
and we assume that there exists $c\in (0,1/3)$ such that
\[s^*\rho(S^*)\leq c.\]
If we have
\[\frac{\sqrt{\tilde\gamma}+\sqrt{\gamma}}{\sqrt{\tilde\gamma}-\sqrt{\gamma}}\leq \frac{1-c}{2c},\]then
\[\P\left\{ \hat S\subset S^*\right\}\geq 1-2M^{1-\gamma/2}.\]
\end{Th}
A similar result was established by \cite{bunea2}  in a slightly less general model. However, her result is based on strong assumptions on the dictionary, namely each function is bounded by a constant $L$ (see Assumption (A2)(a) in \cite{bunea2}). This assumption is mild when considering dictionaries only based on Fourier bases. It is no longer the case when wavelets are considered and Bunea's assumption is satisfied only in the case where $L$ depends on $M$ and $n$ on the one hand and is very large on the other hand. Since $L$ plays a main role in the definition of the tuning parameters of the method, with too rough values for $L$, the procedure cannot achieve satisfying numerical results for moderate values of $n$ even if asymptotic theoretical results of the procedure are good. In the setting of this paper, where we aim at providing calibrated statistical procedures, we avoid such assumptions.

Finally, we have the following corollary.

\begin{cor}\label{oracleadmi3}
We suppose that $A1(s^*)$ is satisfied and that there exists $c\in (0,1/3)$ such that
\[s^*\rho(S^*)\leq c.\] If we have
\[\frac{\sqrt{\tilde\gamma}+\sqrt{\gamma}}{\sqrt{\tilde\gamma}-\sqrt{\gamma}}\leq \frac{1-c}{2c},\] then, with probability
at least $1-4M^{1-\gamma/2}$,
\[\normD{\hat{f}-\fo}^2 \leq \frac{32 s^* \tilde{r}_n^2}{\kappa_{s^*}},
\]
where
$$\tilde{r}_n=\sup_{j=1,\ldots,M}\tilde{r}_{n,j}.$$
\end{cor}

This corollary is a simple consequence of Theorem~\ref{oracleadmi2} with $\lambda=\lambda^*$ and $J_0=S^*$. Taking $\lambda=\lambda^*$ implies that the approximation term vanishes. Taking $J_0=S^*$ implies that the bias term vanishes since the support of the LASSO estimator is included in the the support of $\lambda^*$. In this case, assuming that $\sup_j \|\p_j\|_n<\infty$, the rate of convergence is the classical rate $\frac{s^*\log M}{n}$. 
\section{Estimation algorithm and inferences}\label{algos}
We propose the following estimation procedure for semiparametric estimation of $(\bdtheta, f)$ in model (\ref{mod_vec}), combining the algorithms described in sections \ref{par_est} and \ref{f_alg}:\\

{\bf Estimation Algorithm - ML version:} at iteration $k$,
\begin{itemize}
\item[-] Given the current estimate of $\bdtheta$, $\bdtheta^{(k)}=(\bdbeta^{(k)}, \bdGamma^{(k)}, \sigma^{2(k)})$, and $m$ sampled values of the random effects $\bdphi^{(k,l)}$, $l=1,\ldots,m$, update the estimates of $f$, $f^{(k,l)}$, $l=1,\ldots,m$, with the algorithm described in Section~\ref{f_alg}.
\item[-] Given the current estimates of $f$, $f^{(k,l)}$, $l=1,\ldots,m$, sample $m$ values of the random effects $\bdphi^{(k,l)}$, $l=1,\ldots,m$, and update the value of $\bdtheta$, $\bdtheta^{(k+1)}=(\bdbeta^{(k+1)}, \bdGamma^{(k+1)}, \sigma^{2(k+1)})$ with algorithm (\ref{saem-ml}).
\begin{equation}\label{algML}\end{equation}
\end{itemize}

{\bf Estimation Algorithm - REML version:} at iteration $k$,
\begin{itemize}
\item[-] Given the current estimate of $\tilde{\bdtheta}$, $\tilde{\bdtheta}^{(k)}=(\bdGamma^{(k)}, \sigma^{2(k)})$, and $m$ sampled values of the missing data $\bdz^{(k,l)}=(\bdphi^{(k,l)}, \bdbeta^{(k,l)})$, $l=1,\ldots,m$, update the estimates of $f$, $f^{(k,l)}$, $l=1,\ldots,m$, with the algorithm described in Section~\ref{f_alg}.
\item[-] Given the current estimates of $f$, $f^{(k,l)}$, $l=1,\ldots,m$, sample $m$ values of the missing data $\bdz^{(k+1,l)}=(\bdphi^{(k+1,l)}, \bdbeta^{(k+1,l)})$, $l=1,\ldots,m$, and update the value of $\tilde{\bdtheta}$, $\tilde{\bdtheta}^{(k+1)}=(\bdGamma^{(k+1)}, \sigma^{2(k+1)})$ with algorithm (\ref{saem-reml}).
\begin{equation}\label{algREML}\end{equation}
\end{itemize}

As it is explained in Section \ref{par_est}, for parametric estimation (SAEM or SAEM-REML algorithms alone) the number of chains, $m$, can be set to 1, which still guarantees the convergence towards a local maximum of the log-likelihood. Higher values of $m$, may accelerate the convergence of the algorithms (but in practice, $m$ is always lower than 10).\\
For the global semiparametric estimation procedure, we extend this idea of ``parallel chains'' of values to the estimation of $f$. Indeed, at iteration $k$, the estimation of $f$ depends on the value of the missing data, and thus, from $m$ sampled values $\bdz^{(k,1)},\ldots,\bdz^{(k,m)}$  we obtain $m$ estimates of $f$, $f^{(k,1)},\ldots,f^{(k,m)}$ (see Section \ref{f_est}). Then, in the second step, we use each one of these different estimates of $f$ in parallel to perform parametric estimation (using $f^{(k,l)}$ to sample $\bdz^{(k+1,l)}$ and replacing $\hat{f}_{-}$ by $f^{(k,l)}$ in (\ref{saem-reml}) for the estimation of $\tilde{\bdtheta}$). This is in the case of the REML version of the algorithm, but the same idea underlies the ML version.\\

Inferences on model and individual parameters, $\bdbeta, \bdGamma, \sigma^2$ and $\bdphi$, are performed as in NLMEs (see \cite{Kuhn05} and \cite{Cristian1}). For inferences on the nonlinear function $f$, we propose an empirical approach based on the fact that our algorithm automatically provides large samples of estimates of $f$.\\
Indeed, at each iteration of algorithms (\ref{algML}) and (\ref{algREML}) we obtain $m$ estimates of $f$. The last iterations of the algorithms typically correspond to small values of $\gamma_k$ in algorithms (\ref{saem-ml}) and (\ref{saem-reml}), see Section \ref{applications} for the details. This can be seen as a phase in which the estimates of parameters are stabilized since we assume that convergence has been reached. Let us note by $K$ and $L<K$ the total number of iterations and the number of iterations in the ``stabilization phase'' of the algorithm. Then, by considering the last $L_0 <L$ iterations of the algorithm, we get a large sample of estimates of $f$: $f^{(k,l)}$, $l=1,\dots,m$, $k=L_0+1,\ldots,K$. These $m\times L_0$ estimates of $f$ are obtained conditionally to values of $\bdtheta$ which are supposed to be close to the corresponding ML or REML estimates. Then, we obtain a point estimate for $f$ as:
\begin{equation}\label{f_point_est}
\hat{f}=\dfrac{1}{m\times L_0} \displaystyle \sum_{k=K-L_0+1}^K\sum_{l=1}^m f^{(k,l)}
\end{equation}
and an empirical pointwise $(1-\alpha)100\%$ confidence interval for $f(x)$ as:
\begin{equation*}\label{f_IC}
\left( \hat{f}(x) - z_{\frac{\alpha}{2}} \sqrt{\frac{S^2_{f(x)}}{m\times  L_0}}\,\, , \,\, \hat{f}(x) + z_{\frac{\alpha}{2}} \sqrt{\frac{S^2_{f(x)}}{m\times L_0}} \right),
\end{equation*}
where $S^2_{f(x)}=\dfrac{1}{m\times L_0-1}  \sum_{k=K-L_0+1}^K\sum_{l=1}^m (f^{(k,l)}(x) - \hat{f}(x))^2$ and $z_{\frac{\alpha}{2}}$ is the $1 - \frac{\alpha}{2}$ percentile of a standard normal distribution. This interval is of course not a true $(1-\alpha)100\%$ confidence interval for $f(x)$ but constitutes a good approximation of it. In the same way, we can also construct approximated confidence intervals for the expected response at a given point for a given individual.
This approach is an alternative to the Bayesian confidence intervals proposed by \cite{KeWang}. The idea is similar to bootstrap confidence intervals, with the advantage that here, the samples of estimates are automatically generated by the estimation algorithm.

\section{Application to synthetic and real data}\label{applications}
\subsection{First simulation study: parametric estimation}\label{simulation-study}
As a first step, we want to validate through simulation our parametric estimation strategy alone, based on the SAEM algorithm, and to compare it, in the framework of SNMMs, to the approximate method {\it nlme} of \cite{KeWang}. In order to be able to asses only the differences induced by the use of different parametric estimation algorithms, we will use the same nonparametric estimation algorithm for the estimation of $f$, namely the procedure proposed by \cite{KeWang}. In the next section we will compare the whole versions, including nonparametric estimation, of both approaches.\\
To this end, we realized the following simulation study. As in Example~1, data were generated from the model:
$$y_{ij}=\phi_{1i}+\exp (\phi_{2i}) 2 f\left(\frac{j}{N}-\frac{\exp (\phi_{3i})}{1+\exp (\phi_{3i})}\right) + \varepsilon_{ij},\quad i=1,\cdots,N, \,\, j=1,\cdots,J, $$
 where $\varepsilon_{ij}\sim \mathcal{N}(0,\sigma^2)$ and $\phi_i=(\phi_{1i},\phi_{2i},\phi_{3i})^T\sim \mathcal{N}(\mu,\Gamma)$ with $\mu=(\mu_1,\mu_2,\mu_3)^T$.  Here, the nonlinear function was set to $f(t)=\sin(2\pi t)$.
The following parameter values were used for simulation:\\

$N=J=10$, $\mu=(1,0,0)^T$, $\sigma^2=1$ and $\Gamma$ is diagonal with $diag(\Gamma)=(1,0.25,0.16)$.\\

\noindent These data were analyzed using two semiparametric procedures: our SAEM based method combined with the nonparametric algorithm of Ke and Wang's (called {\it semi-SAEM}) and Ke and Wang's procedure for semiparametric models (called {\it snm}).
For the SAEM algorithm, we used 80 iterations and the following sequence $(\gamma_k)$: $\gamma_k=1$ for $1\leq k\leq 50$ and $\gamma_k=1/(k-50)$ for $51\leq k \leq 80$. We also considered $m=5$ chains in each iteration. For the nonparametric estimation algorithm common to both procedures, following \cite{KeWang} we considered that $f$ is periodic with period equal to 1 and $\int_{0}^1 f=0$, i.e. $f\in W^0_2(per)=W_2(per)\ominus span\{1\}$ where $W_2(per)$ is the periodic Sobolev space of order $2$ in $L^2$ and $span\{1\}$ represents the set of constant functions.\\ The same initial values were used for both methods:\\

$\mu_0=(1,0,0)$, $\sigma^2_0=2$ and $\diag (\Gamma_0)=(\gamma_1^0,\gamma_2^0,\gamma_3^0)=(1,0.3,0.1).$ \\

Tables~\ref{tmedias} and \ref{tvar} sumarize the performance of both methods over 100 simulated data sets. For each parameter we show the sample mean, the mean squared error ($MSE(\hat{\theta})=\dfrac{1}{100}\sum_{i=1}^n (\theta-\hat{\theta}_i)^2$), and a 95\% confidence interval computed over the total number of simulations.

\begin{table}
\begin{center}
\begin{tabular}{ccccc}
& Method & $\mu_1$ & $\mu_2$ & $\mu_3$\\\hline
True Value & & 1 & 0 & 0 \\\hline
Mean & semi-SAEM &1.06 & 0.31 & 0.27\\
& snm &  1.05 & 0.26 & -0.01\\
MSE & semi-SAEM & 0.12 & 0.16 & 0.10\\
& snm &  0.12 & 0.11 & 0.01\\
95 \% C.I. & semi-SAEM & [0.99;1.12] & [0.27;0.36] & [0.23;0.30]\\
 & snm & [0.99;1.12] & [0.22;0.30] & [-0.02;0.01]\\\hline
\end{tabular}
\caption{ML procedure: Mean, MSE and 95\% confidence interval of mean components.}\label{tmedias}
\end{center}
\end{table}

\begin{table}
\begin{center}
\begin{tabular}{cccccc}
& Method & $\gamma_1$ & $\gamma_2$ & $\gamma_3$ & $\sigma^2$\\\hline
True Value & & 1 & 0.25 & 0.16 & 1 \\\hline
Mean & semi-SAEM & 0.86 & 0.24 & 0.16 & 0.95\\
& snm &  0.89 & 0.19 & 0.14 & 0.99\\
MSE & semi-SAEM & 0.22 & 0.02 & 0.01 & 0.03\\
& snm &   0.22 & 0.02 & 0.01 & 0.03\\
95 \% C.I. & semi-SAEM & [0.77;0.95] & [0.21;0.27] & [0.14;0.17] & [0.92;0.98]\\
 & snm &  [0.80;0.98] & [0.17;0.21] & [0.13;0.16] & [0.96;1.02]\\\hline
\end{tabular}
\caption{ML procedure: Mean, MSE and 95\% confidence interval of variance components obtained with {\it semi-SAEM} and {\it snm}.}\label{tvar}
\end{center}
\end{table}

We also compared the REML estimates obtained with our method and with {\it snm} (using the REML version of {\it nlme}) for the same simulated data sets. The results are summarized in Table \ref{tvar.REML}. It can be seen that the mean values for the REML estimates obtained with both procedures were closer to the simulated values, especially for parameters $\gamma_1$. Moreover, the individual confidence intervals of REML estimates of this parameter, at a 95\% level, include the true value for these parameters on the contrary to the ML estimates, showing that REML versions of the  algorithms were able to correct the bias observed with ML.
If we compare our method and {\it snm}, for both procedures ML and REML, we obtained  results that are similar but it seems that our REML estimates are closer to the simulated values than those obtained with Ke and Wang's method.

An important issue to discuss is the convergence of estimates with this kind of iterative maximization algorithms. It is well known that approximate methods for maximum likelihood estimation often present numerical problems and even fail to converge in the framework of NLME estimation (see \citep{nlme_pb} for instance). An advantage of the exact likelihood method is exactly to avoid those convergence problems as it was established by \cite{Kuhn05}. In this simulation study, we have to say that both {\it semi-SAEM} and {\it snm} achieved convergence for all the data sets. However, we also tried to fit a nonlinear mixed effects model to the simulated data, that is, assuming that $f$ was known and estimating only the fixed and random effects with {\it SAEM} and {\it nlme}, and in that case the second algorithm failed to converge for several data sets. It seems that in this case the combination of {\it nlme} with a nonparametric algorithm to perform semiparametric estimation solves the numerical problems encountered by {\it nlme} on its own. However, this is not true in general as we will see in the next simulation study.

\begin{table}
\begin{center}
\begin{tabular}{cccccc}
& Method & $\gamma_1$ & $\gamma_2$ & $\gamma_3$ & $\sigma^2$\\\hline
True Value & & 1 & 0.25 & 0.16 & 1 \\\hline
Mean & semi-SAEM & 0.99 & 0.25 & 0.16 & 0.95\\
& snm &  0.92 & 0.19 & 0.15 & 1.02\\
MSE & semi-SAEM & 0.21 & 0.03 & 0.01 & 0.03\\
& snm &   0.23 & 0.02 & 0.01 & 0.03\\
95 \% C.I. & semi-SAEM & [0.89;1.08] & [0.22;0.28] & [0.14;0.18] & [0.92;0.98]\\
 & snm &  [0.83;1.02] & [0.17;0.22] & [0.13;0.17] & [0.98;1.05]\\\hline
\end{tabular}
\caption{REML procedure: Mean, MSE and 95\% confidence interval of variance components obtained with {\it semi-SAEM} and {\it snm}.}\label{tvar.REML}
\end{center}
\end{table}


\subsection{Second simulation study: semiparametric estimation}
\begin{figure}[!htp]
 \centering
 \includegraphics[width=12cm,height=12cm]{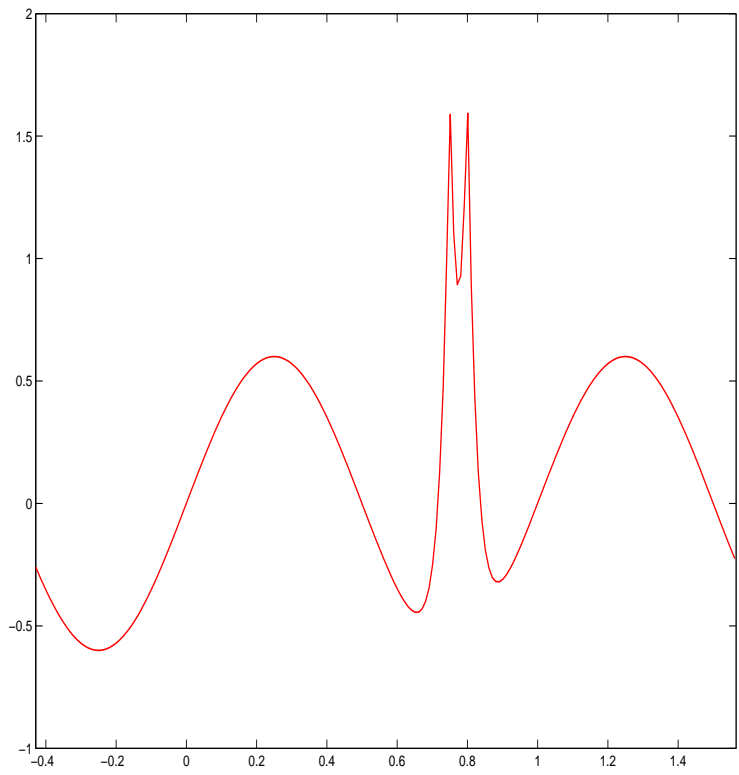}
 \caption{Real function $f$ in the semiparametric simulation study.}\label{fig.real.f}
\end{figure}

In order to test our LASSO-based estimator, we modified the model introduced in section \ref{simulation-study} as follows:
$$y_{ij}=\phi_{1i}+\exp (\phi_{2i}) 2 f\left(\frac{j}{N}-\frac{\exp (\phi_{3i})}{1+\exp (\phi_{3i})}\right)+ \varepsilon_{ij},\quad i=1,\cdots,N, \, \, j=1,\cdots,J, $$
where
$\varepsilon_{ij}\sim \mathcal{N}(0,\sigma^2)$ and $\phi_i=(\phi_{1i},\phi_{2i},\phi_{3i})^T\sim \mathcal{N}(\mu,\Gamma)$ with $\mu=(\mu_1,\mu_2,\mu_3)^T$. Here, $f(\cdot)$ is a mixture of one trigonometric function and two Laplace densities (see Figure~\ref{fig.real.f}).
$$f(t)=0.6\times\sin(2\pi t) + 0.2\times \left( \frac{e^{-40|t-0.75|}}{2\times\int_0^1 e^{-40|t-0.75|}} \right)+ 0.2\times \left( \frac{e^{-40|t-0.8|}}{2\times \int_0^1 e^{-40|t-0.80|}} \right).$$

Data were simulated using the following parameters:\\

$N=10$, $J=20$, $\mu=(1,0,0)^T$, $\sigma^2=0.4$ and $\Gamma$ is diagonal with $diag(\Gamma)=(0.25,0.16,0.04)$.\\

\noindent Now data were analyzed using the two following semiparametric procedures: our SAEM and LASSO based method (called {\it LASSO-SAEM}) and Ke and Wang's procedure for semiparametric models, still denoted {\it snm}. For both methods we obtained the REML estimates of parameters.\\
It is necessary to specify several values in order to run our algorithm, such as the choice of the LASSO's tunning parameter $\gamma$ and the inputs of the SAEM algorithm (initial values, step sizes $\gamma_k$, number of chains in the MCMC step, number of burn-in iterations, and total number of iterations).
For the latter, we used again 80 iterations with $\gamma_k=1$ for $ 1\leq k\leq 50$ and $\gamma_k=1/(k-50)$ for $51\leq k \leq 80$, and we considered $m=5$ chains in each iteration.
The initial values, which were also used with \emph{snm}, were:\\

\begin{figure}[!htp]
 \centering
 \includegraphics[width=12cm,height=12cm]{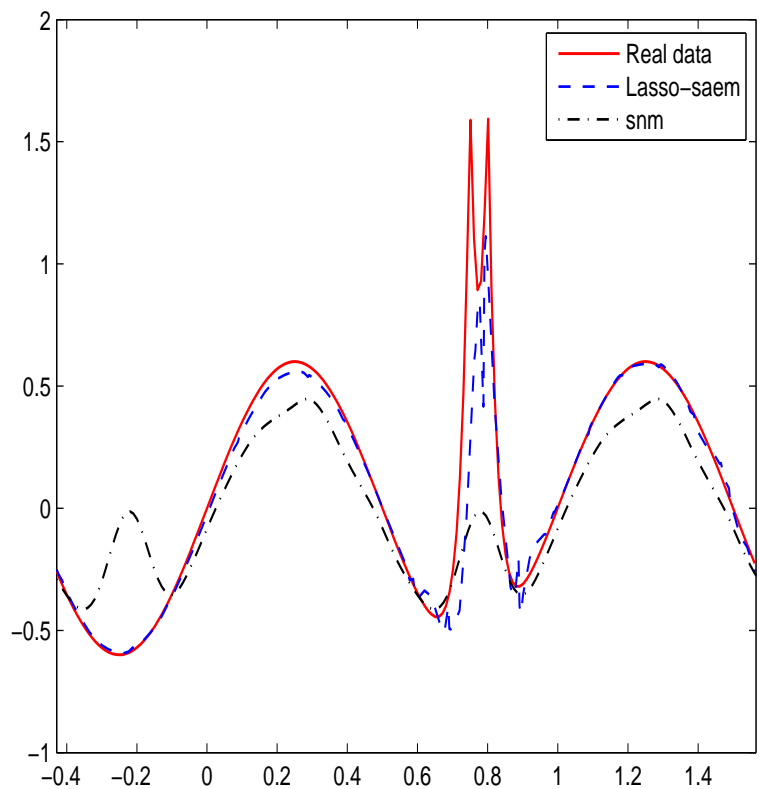}
 \caption{Real function $f$ (solid line) and its estimates obtained with {\it LASSO-SAEM} (dashed line) and {\it snm} (dash-dotted line) for a particular data set in the semiparametric simulation study.}\label{fig.data.panel}
\end{figure}

\begin{figure}[!htp]
 \centering
 \includegraphics[width=16cm,height=16cm,angle=-90]{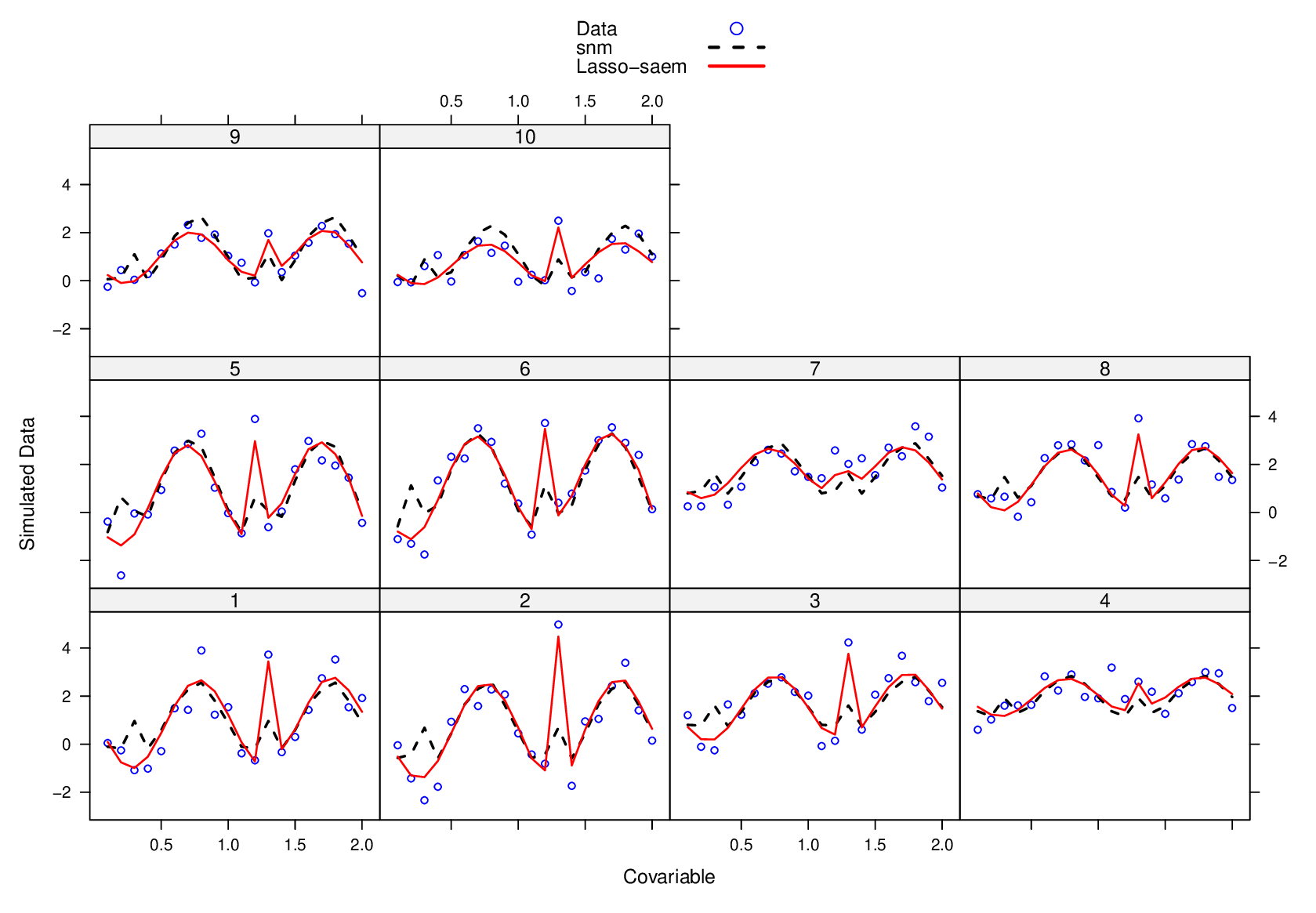}
 \caption{Simulated data and fitted curves obtained with {\it LASSO-SAEM} (solid line) and {\it snm} (dashed line) for a particular data set in the semiparametric simulation study.}\label{fig.data}
\end{figure}

$\mu_0=(1,0,0)$, $\sigma^2_0=2$ and $\diag (\Gamma_0)=(\gamma_1^0,\gamma_2^0,\gamma_3^0)=(1,0.3,0.1).$ \\
The nonparametric LASSO step has been performed with $\gamma=1/3$. Larger values of $\gamma$ did not allow, for some datasets, to stabilizing the convergence of some parameters, in particular the variance $\gamma_2$, and smaller values of $\gamma$ provided similar results to the one presented here.
The dictionary chosen combined very different orthonormal families, namely Fourier functions with Haar wavelets, which ensured a sufficiently incoherent design in the spirit of Section~\ref{f_est}. More precisely, our dictionary was composed by the following Fourier functions $\{t\mapsto 1; t\mapsto\cos(\pi t);t\mapsto \sin(\pi t);t\mapsto\cos(2\pi jt), t\mapsto \sin(2\pi jt), j= 1,\cdots,5\}$ and by the Haar wavelet basis with resolution between $2^4$ and $2^7$, with a total size of 245 functions. Note that the data $\tilde{\bdx}_{ij}=c(\bdphi_i;{\bdx}_{ij})$ belongs approximately to $[-0.4,1.6]$. For {\it snm}, it seemed reasonable to consider that $f\in W^0_2(per)$ since if we look at a simulated data set (see Figure \ref{fig.data} for example), we can see clearly the periodic structure in the data.\\
In Figures \ref{fig.data.panel} and \ref{fig.data}, we can see the estimates of $f$ and the fitted data with the two methods for a specific simulated data set.\\

Results for REML estimates obtained with {\it LASSO-SAEM} and {\it snm} for 100 simulated data sets are summarized in Table \ref{tvar.REML-LASSO}. We can see that the means of the estimates obtained with our method are close to their real values except for the variance of the error, $\sigma^2$, since our method tends to overestimate that parameter. However, we get overall better results than using the {\it snm} methodology (except for $\gamma_1$).

\begin{table}[!h]
\begin{center}
\begin{tabular}{cccccc}
& Method & $\gamma_1$ & $\gamma_2$ & $\gamma_3$ & $\sigma^2$\\\hline
True Value & & 0.25 & 0.16 & 0.04 & 0.4 \\\hline
Mean & LASSO-SAEM & 0.18 & 0.14 & 0.03 & 0.69\\
& snm &  0.21 & 0.11 & 0.03 & 0.90\\
MSE & LASSO-SAEM & 0.01 & 0.01 & 4.0e-4 & 0.12\\
& snm &   0.02 & 0.01 & 5.9e-4 & 0.27\\
95 \% C.I. & LASSO-SAEM & [0.16;0.20] & [0.12;0.15] & [0.030;0.037] & [0.66;0.73]\\
 & snm &  [0.18;0.25] & [0.09;0.14] & [0.028;0.042] & [0.86;0.94]\\\hline
\end{tabular}
\caption{REML procedure: Mean, MSE and 95\% confidence interval of variance components obtained with {\it LASSO-SAEM} and {\it snm}.}\label{tvar.REML-LASSO}
\end{center}
\end{table}
\medskip

\begin{figure}[!htp]
 \centering
 \includegraphics[width=15cm,height=15cm]{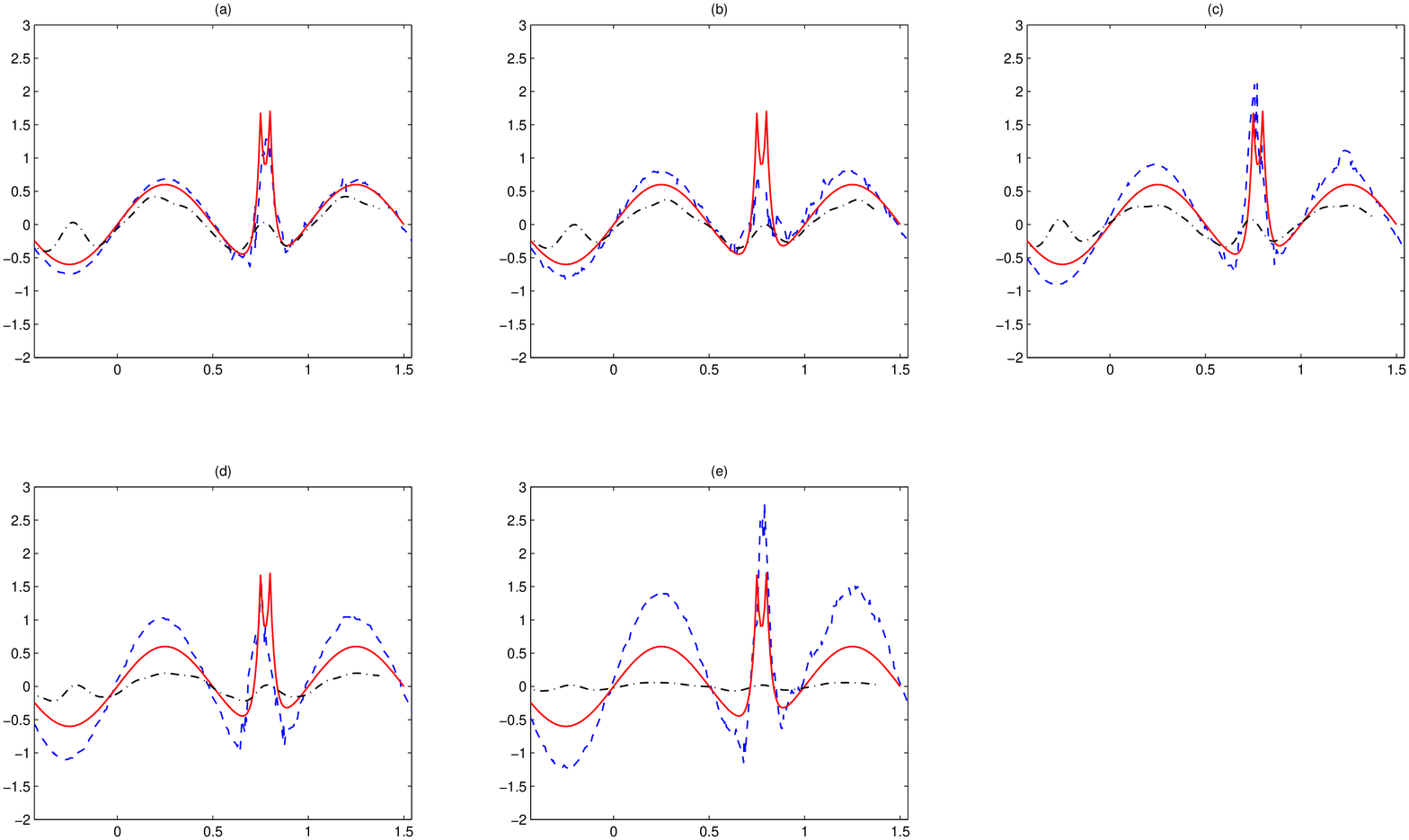}
 \caption{Estimated functions corresponding to the five quantiles of ISE ((a) minimum, (b) 1/4 quantile, (c) median, (d) 3/4 quantile and (e) maximum) obtained with {\it LASSO-SAEM} (dashed line) and {\it snm} (dash-dotted line) compared to the true function $f$ (solid line) for the total of the 100 simulated data sets in the semiparametric simulation study.}\label{fig.ISE}
\end{figure}

An important issue for this kind of problem is the estimation of the nonlinear function $f$. Then, to evaluate the accuracy of the estimation, we calculated the Integrated Square Error (ISE) of $\hat{f}$ for each simulated data set. Figure \ref{fig.ISE} provides a summary of estimates of $f$ using {\it LASSO-SAEM} and {\it snm}. We computed the ISE for each estimate of $f$ and plotted the estimates corresponding to (a) the minimum, (b) 1/4 quantile, (c) median, (d) 3/4 quantile and (e) maximum ISEs. We can see that our method outperforms {\it snm} in the estimation of $f$, in the sense that our estimates are able to detect the presence of the peaks in the original function.\\
As for the functions of the dictionary selected with our LASSO method, it is interesting to note that the 100 linear combinations of functions of the dictionary obtained for each one of the 100 data sets have a length which varies between 10 and 32 functions, with an average length equal to 20. Furthermore, in 98\% of the cases, the method selects the function $\sin (2\pi t)$ with the highest coefficient. For the remaining two data sets,  the functions $\sin (6\pi t)$ and $\sin (10\pi t)$ are selected. For all the replicates, in addition to these sine functions, the rest of the selected functions are related to the Haar wavelets with smaller coefficients.

It is very important to point out that the results obtained with {\it snm} are based only on 51 data sets since the function did not reach convergence in 46 data sets and in other 3 data sets we obtained incoherent estimation of the nonlinear function, when using the default setup of the $snm$ algorithm (REML estimation and Generalized Cross Validation for the choice of the penalized parameter).  By contrast, our method achieved convergence for all simulated data sets with the specific setup used here (choice of $\gamma$, initial values, number of chains, step sizes $\gamma_k$, number of iterations, etc $\ldots$).

\subsection{Application to on-line auction data}
\begin{figure}[!htp]
 \centering
 \includegraphics[trim = 15mm 5mm 10mm 0mm, clip,width=16.5cm,height=9cm]{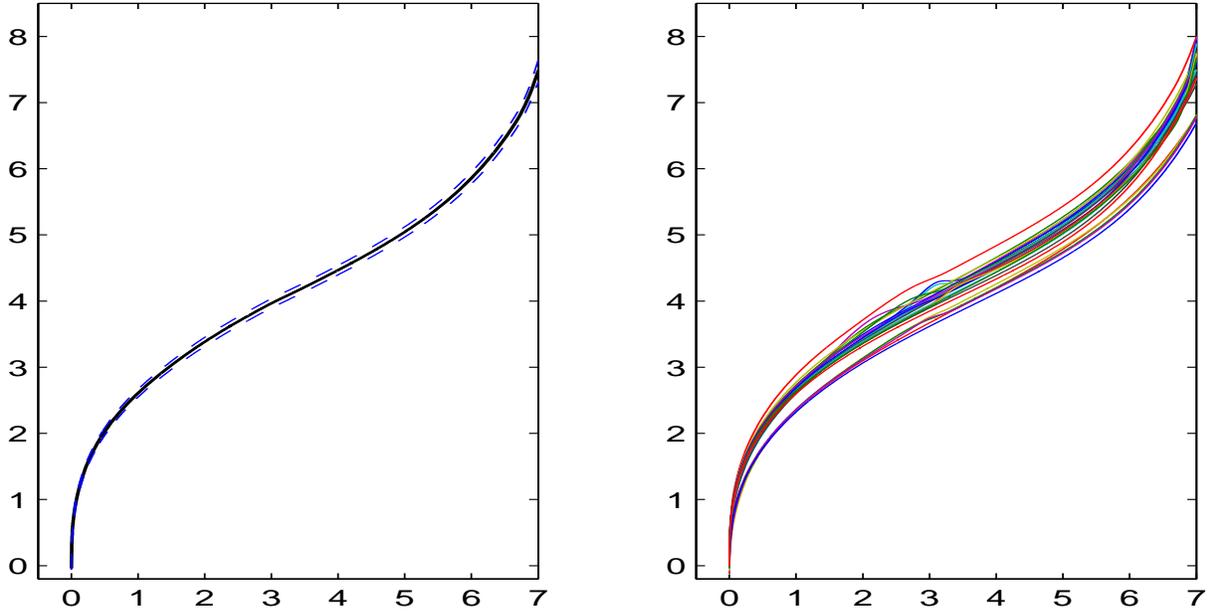}
 \caption{Left: Estimated nonlinear function $\hat{f}$ (solid line) and $95\%$ confidence band (dashed lines) in the on-line auction dataset. Right: Last 24 LASSO estimates whose empirical mean provides $\hat{f}$.}\label{AuctionFF}
\end{figure}

\begin{figure}[!htp]
 \centering
 \includegraphics[trim = 5mm 62mm 5mm 62mm, clip,width=20cm,angle=-90]{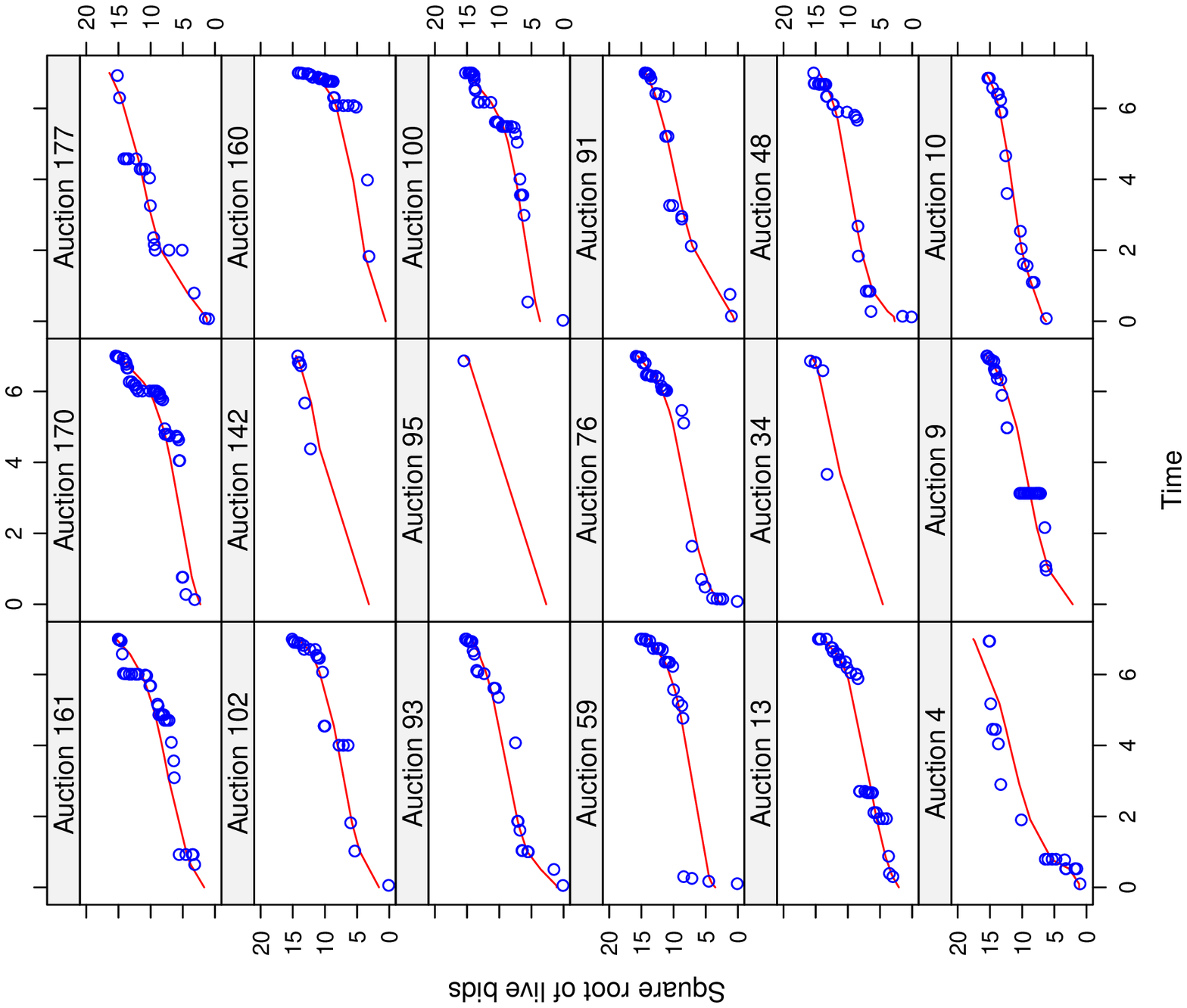}
 \caption{Observed live bids (circles) and fitted price curves (solid lines) for a subset of 18 auctions.}\label{AuctionF2}
\end{figure}

Modelling of price paths in on-line auction data has received a lot of attention in the last years \citep{AD1, AD2, AD3, AD4}. One of the reasons is the availability of huge amounts of data made public by the on-line auction and shopping website eBay.com, which has become a global market place in which millions of people worldwide buy and sell products. The price evolution during an auction can be thought as a continuous process which is observed discretely and sparsely only at the instants in which bids are placed. In fact, bids tend to concentrate at the beginning and at the end of the auction, responding to two typically observed phenomena, ``early bidding'' and ``bid sniping'' (a situation in which ``snipers'' place their bids at the very last moment).\\
To our knowledge, \cite{AD_mm} provide the first attempt to model price paths taking into account the dependence among different auctions. This is an important consideration, since in practice bidders can participate in different auctions that take place simultaneously. They propose a semiparametric additive mixed model with a boosting estimation approach. In the same line, but considering a more complex interaction of the random effects and the unknown nonlinear function, we propose the following shape-invariant model for the price paths:
$$y_{ij}=\phi_{1i}+\exp(\phi_{2i})f (t_{ij} - \phi_{3i} )+ \varepsilon_{ij},\quad i=1,\cdots,N, \,\, j=1,\cdots,n_i, $$
where $\varepsilon_{ij}\sim \mathcal{N}(0,\sigma^2)$ and $\phi_i=(\phi_{1i},\phi_{2i},\phi_{3i})^T\sim \mathcal{N}(\mu,\Gamma)$ with $\mu=(\mu_1,\mu_2,\mu_3)^T$. We introduce an individual random horizontal shift, $\phi_{3i}$, to model the possible delay of the price dynamics in some auctions with respect to the rest.\\
We analyzed a set of 183 eBay auctions for Palm M515 Personal Digital Assistants (PDA), of a fixed duration of seven days, that took place between March and May, 2003. This is the dataset used in \cite{AD_mm} and it is publicly available at\\ \texttt{http://www.rhsmith.umd.edu/digits/statistics/data.aspx}. We were interested in modelling the live bids, that is, the actual prices that are shown by eBay during the live auction. Note that these are different from the bids placed by bidders during the auction, which are the prices recorded in the bid history published by eBay after the auction closes. Then, a transformation on the bid records is required to recover the live bids (see \cite{AD1} for details).\\
The live bids range from \$0.01 to \$300 and form a sequence of non decreasing prices for each auction. We typically observe between 10 and 30 bids per auction, although there are auctions with only two bids. We have a total of 3280 bids for the 183 auctions. Following \cite{AD_mm}, we considered the square root of live bids to reduce the price variability. We run the REML version of our \emph{LASSO-SAEM} algorithm, of which we performed 100 iterations with the following sequence of decreasing steps $(\gamma_k)$: $\gamma_k=1$ for $1\leq k\leq 60$ and $\gamma_k=1/(k-60)$ for $61\leq k \leq 100$. We also considered $m=3$ chains in each iteration. The dictionary for nonparametric estimation was composed by a combination of B-splines of degrees three and four, with 17 knots unequally spaced so that most of the knots were in those places with more data observed (at the beginning, at the end and at the middle of the interval), 10 power functions, 10 exponential functions and 5 logit functions, with a total size of 64 functions. The estimate of $f$ is monotone, as expected by the nature of the data, and presents two steepest parts at the beginning and at the end of the interval. At each iteration of the algorithm the estimated function at the nonparametric step is a sparse combination of the functions of the dictionary. In fact, the set of functions selected by the LASSO method at the last iterations of the algorithm is almost constant, containing mainly two functions, $\varphi(x)=x^{0.35}$ and $\varphi(x)=\exp(0.9\,x)$, and in some iterations a small component of a cubic B-spline around the middle of the interval. In Figure \ref{AuctionFF} we present the last 24 estimates $f^{(k,l)}$ from which we have obtained $\hat{f}$ as in (\ref{f_point_est}), and $\hat{f}$, together with a 95\% pointwise confidence band. These results have been obtained with $\gamma=2$ as the value for the tunning parameter in the LASSO estimation step.\\
The estimates for $\mu$ and $\Gamma$ are presented in Table~\ref{AuctionT1}. In Figure \ref{AuctionF2} we present the observed live bids and the model fit for 18 chosen auctions with different price profiles. We can appreciate how the fitted model provides in general an accurate fit of the final price, even in the cases when ``bid sniping'' is present.

\begin{table}
\begin{center}
\begin{tabular}{cccc|c}
           & $\phi_1$ & $\phi_2$ & $\phi_3$ & \\\hline
Mean       &   1.04       & 0.18          &  -0.06       & \\\hline
Correlation& 1 (7.68)        &   -0.02       &  0.41        & $\phi_1$\\
Matrix     & -0.02  &       1  (0.19) &0.37          & $\phi_2$\\
(variances)& 0.41   &   0.37        &   1 (0.23)      & $\phi_3$\\ \hline
$\sigma^2$ &          &    1.93    &         & \\\hline
\end{tabular}
\caption{Estimated mean vector and covariance matrix of the random effects and estimated error variance in the on-line auction dataset.}\label{AuctionT1}
\end{center}
\end{table}

\section{Conclusions and discussion}\label{con}

Semiparametric nonlinear mixed effects models cover a wide range of situations and generalize a large class of models, such as nonlinear mixed effects models or self-modelling nonlinear regression models among others. We have proposed a new approach for estimation in SNMMs combining an exact likelihood estimation algorithm with a LASSO-type procedure. Our strategy relies on an iterative procedure to estimate $\bdtheta$ conditioned on $f$ and vice versa, which allow us to tackle the parametric and the nonparametric problem independently. This makes possible the use of fast algorithms providing an accurate and computationally efficient estimation method.\\
Concerning parametric estimation, our simulation results illustrate our method and point out some important advantages of using an exact likelihood estimation algorithm instead of likelihood approximation methods, such as convergence of the estimates. The REML version of our algorithm, corrects the estimation of variance components accounting for the loss of degrees of freedom from estimating the fixed effects and provide satisfactory results. However, as it was already pointed out in the comments to \cite{KeWang}, it will be important to define a REML estimator that can also take into account the loss of degrees of freedom from estimating the nonlinear function $f$.

As for nonparametric estimation, the dictionary approach allows us to obtain interesting interpretation with respect to the functions of the dictionary selected by the procedure. For instance, we can detect trends, frequencies of sinusoids or location and heights of peaks of the common shape represented by the estimated function $f$. We have observed that our LASSO estimate achieves good theoretical and numerical results if the dictionary is wealthy and incoherent enough. From the theoretical point of view, incoherence is expressed, in this paper, by Assumption $\mbox{A1}(s)$  or by the quantity $\rho(S^*)$ defined in Section~\ref{supp-sect}. These incoherence assumptions are hard to check in practice and we do not know if they can be relaxed in our setting. We mention that the method is quite sensitive to the choice of the dictionary. Indeed, in our application to on-line auction data we have detected that differences in the size of the dictionary, but not necessarily in the nature of the function families therein included, may lead to slightly different estimated functions, in the sense that we may obtain rougher and smoother versions of a similar function.\\
In Section~\ref{f_est}, the particular structure of the observations (where we have $n_i$ observations for each individual $i$) is not used for applying the standard LASSO-procedure. But a natural and possible extension of this work would be to take into account this structure and then to apply a more sophisticated LASSO-type procedure inspired, for instance, by the group-LASSO proposed by \cite{YL06} to achieve better results. This is a challenging research axis we wish to investigate from a theoretical and practical point of view.

Among other possible extensions of this work, a very promising one would be the use of the nonparametric techniques herein described for density estimation (in the spirit of \citep{berlepenrivo}) of the ramdom errors, assuming that they do not need to be normal. Indeed, the recent work of \cite{Adeline} deals with this problem in the case of a linear mixed effects model. Its generalization to NLMEs or even SNMMs is a real challenge.

\appendix
\section*{Appendix. The proofs}
\subsection*{Preliminary lemma}
\begin{lemma}\label{prel}
For $1\leq j\leq M$, we consider the event $\mathcal{A}_j=\left\{|V_j|< r_{n,j}\right\}$ where $V_j=\frac{1}{n}\sum_{i=1}^n b_i \p_j(x_i)\e_i$. Then,
\[\P\left(\mathcal{A}_j\right)\geq 1-M^{-\gamma/2}.\]
\end{lemma}
{\bf Proof of Lemma~\ref{prel}:}
 We have
\begin{align*}
\P\left(\mathcal{A}_j^c\right)&\leq  \P\left(\sqrt{n}|V_j|/(\sigma\|\p_j\|_n)\ge \sqrt{n}r_{n,j}/(\sigma\|\p_j\|_n)\right)\\
& \le \P\left(|Z|\geq \sqrt{\gamma\log M}\right)\\
& \le M^{-\gamma/2}
\end{align*}
where $Z$ is a standard normal variable.\hfill$\square$
\subsection*{Proof of Theorem~\ref{oracleadmi}}
Let $\lambda\in\R^M$ and $J_0$ such that $|J_0|=s$. We have
\begin{align*}
\|f_\lambda-\fo\|_n^2=\|\hat{f}-\fo\|_n^2+\|f_\lambda-\hat{f}\|_n^2+\frac{2}{n}\sum_{i=1}^nb_i^2\left(\hat{f}(x_i)-\fo(x_i)\right)\left(f_\lambda(x_i)-\hat{f}(x_i)\right).
\end{align*}
We have $\|f_\lambda-\hat{f}\|_n^2=\|f_{\Delta}\|_n^2$ where $\Delta=\lambda-\hat{\lambda}$. Moreover
$$A=\frac{2}{n}\sum_{i=1}^nb_i^2\left(\hat{f}(x_i)-\fo(x_i)\right)\left(f_\lambda(x_i)-\hat{f}(x_i)\right)=2\sum_{j=1}^M(\lambda_j-\hat{\lambda}_j)\left[(G\hat{\lambda})_j-\beta_j\right],$$
where
$$\beta_j=\frac{1}{n}\sum_{i=1}^nb_i^2\p_j(x_i)f(x_i).$$
Since $\hat{\lambda}$ satisfies the Dantzig constraint, we have with probability at least $1-M^{1-\gamma/2}$, for any $j\in\{1,\ldots,M\}$,
$$|(G\hat{\lambda})_j-\beta_j|\leq |(G\hat{\lambda})_j-\hat\beta_j|+|\hat\beta_j-\beta_j|\leq 2r_{n,j}$$ and
$|A|\le 4r_n\|\Delta\|_1$.
This implies that
\begin{align*}
\|\hat{f}-\fo\|_n^2\le \|f_\lambda-\fo\|_n^2+4r_n\|\Delta\|_1-\|f_\Delta\|_n^2.
\end{align*}
Moreover using Lemma 1 and Proposition 1 of \cite{berlepenrivo} (where the norm $\|\cdot\|_2$ is replaced by $\|\cdot\|_n$), we obtain that
\begin{equation}\label{relationdelta}\left( \normu{\Delta_{\JoC}} - \normu{\Delta_{\Jo}} \right)_+
\leq
2\normu{\lambda_{\JoC}}
+ \left(\normu{\hat\la} -  \normu{\la} \right)_+
\end{equation}
and
\begin{align*}
 \normD{f_\Delta} &\geq \kappa_{s}\normd{\Delta_{\Jo}} - \frac{\mu_{s}}{\sqrt{|\Jo|}}
 \left( \normu{\Delta_{\Comp{\Jo}}}-\normu{\Delta_{\Jo}}\right)_+\\
& \geq \kappa_{s}\normd{\Delta_{\Jo}} - 2 \frac{\mu_{s}}{\sqrt{|\Jo|}}\Lam.
\end{align*}
Note that Proposition 1 of \cite{berlepenrivo} is obtained using Lemma 2 and Lemma 3 of \cite{berlepenrivo}. In our context, Lemma 2 and Lemma 3 can be proved in the same way by replacing the norm  $\|\cdot\|_2$ by $\|\cdot\|_n$ and by considering $P_{J_{01}}$ as the projector on the linear space spanned by $(\p_j(x_1),\ldots,\p_j(x_n))_{j\in J_{01}}$.

Now following the same lines as Theorem 2 of \cite{berlepenrivo}, replacing $\kappa_{J_0}$ by $\kappa_s$ and $\mu_{J_0}$ by $\mu_s$, we obtain the result of the theorem.
\subsection*{Proof of Theorem~\ref{oracleadmi2}}
We consider $\hat\lambda^{D}$ defined by
\[
\hat\lambda^{D}=\argmin_{\lambda\in\R^M} \normu{\lambda}
\quad\mbox{such that $\lambda$ satisfies
the Dantzig
constraint (\ref{dantzig})}.\]
Denote by $\hat{f}^D$ the estimator $f_{\hat{\lambda}^D}$.
Following the same lines as in the proof of Theorem \ref{oracleadmi}, it can be obtained that, with probability
at least $1-M^{1-\gamma/2}$,
for any  integer $s<n/2$  such
that (\ref{Ass1}) holds,
 we have for any $\betamod>0$,
\[\normD{\hat{f}^D-\fo}^2 \leq \inf_{\lambda\in\R^M}
\inf_{\substack{J_0 \subset \{1,\ldots,M\}\\
  |J_0|=s}} \left\{ \normD{f_\lambda-\fo}^2+\betamod \left(1+\frac{2\mu_s}{\kappa_s}\right)^2\frac{
\Lam^2}{s}+16s
\left(\frac{1}{\betamod}+\frac{1}{\kappa_s^2}\right)
r_n^2 \right\},
\]
where here
\[
\Lam = \normu{\lambda_{\Jo^C}}+\frac{\left(\normu{\hat\lambda^{D}}-\normu{\lambda}\right)_+}{2}.
\]
If the infimum is only taken over the vectors $\lambda$ that satisfy the Dantzig constraint, then, with the same probability we have
\begin{equation}\label{prop1}\normD{\hat{f}^D-\fo}^2 \leq \inf_{\lambda\in\mathcal{D}}
\inf_{\substack{J_0 \subset \{1,\ldots,M\}\\
  |J_0|=s}} \left\{ \normD{f_\lambda-\fo}^2+\betamod \left(1+\frac{2\mu_s}{\kappa_s}\right)^2\frac{
\|\lambda_{J_0^C}\|_{l_1}^2}{s}+16s
\left(\frac{1}{\betamod}+\frac{1}{\kappa_s^2}\right)
r_n^2 \right\}.
\end{equation}
Following the same lines as the proof of Theorem \ref{oracleadmi}, replacing $\lambda$ by $\hat{\lambda}^D$, we obtain, with probability
at least $1-M^{1-\gamma/2}$,
\begin{align*}
\|\hat{f}-\fo\|_n^2\le \|\hat{f}^D-\fo\|_n^2+4r_n\|\Delta\|_1-\|f_\Delta\|_n^2,
\end{align*}
with $\Delta=\hat{\lambda}-\hat{\lambda}^D$. Applying
(\ref{relationdelta}) where $\hat{\lambda}$ plays the role of $\lambda$
and $\hat{\lambda}^D$ the role of $\hat{\lambda}$, the vector $\Delta$ satisfies \begin{equation*}\left( \normu{\Delta_{\JoC}} - \normu{\Delta_{\Jo}} \right)_+
\leq
2\normu{\hat{\lambda}_{\JoC}}.
\end{equation*}
Following the same lines as in the proof of Theorem \ref{oracleadmi}, we obtain that for each $J_0 \subset \{1,\ldots,M\}$
such that $|J_0|=s$
\begin{equation}\label{prop2}\normD{\hat{f}-\fo}^2 \leq
\left\{ \normD{\hat{f}^D-\fo}^2+\betamod \left(1+\frac{2\mu_s}{\kappa_s}\right)^2\frac{
\|\hat{\lambda}_{J_0^C}\|_{l_1}^2}{s}+16s
\left(\frac{1}{\betamod}+\frac{1}{\kappa_s^2}\right)
r_n^2 \right\}.
\end{equation}
Finally, (\ref{prop1}) and (\ref{prop2}) imply the theorem.

\subsection*{Proof of Theorem~\ref{theosupport}}
We first state the following lemma.
\begin{lemma}\label{unicite}
We have for any $u\in\R^M$,
\[\mbox{crit}(\hat\la+u)- \mbox{crit}(\hat\la)\geq \left\|\sum_{k=1}^Mu_k\p_k\right\|_n^2.\]
\end{lemma}
{\bf Proof of Lemma~\ref{unicite}:}
Since for any $\la$,
\[\mbox{crit} (\lambda)= \frac{1}{n} \sum_{i=1}^n \left(y_i-b_i f_\lambda(x_i)\right)^2+2\sum_{j=1}^M \tilde r_{n,j}|\lambda_j|,\]
\begin{eqnarray*}
\mbox{crit}(\hat\la+u)- \mbox{crit}(\hat\la)&=&
\frac{1}{n} \sum_{i=1}^n \left(y_i-b_i\sum_{k=1}^M\hat\la_k\p_k(x_i)-b_i\sum_{k=1}^Mu_k\p_k(x_i)\right)^2
+2\sum_{j=1}^M \tilde r_{n,j}|\hat\lambda_j+u_j|
\\&&\hspace{1cm}-\frac{1}{n} \sum_{i=1}^n \left(y_i-b_i\sum_{k=1}^M\hat\la_k\p_k(x_i)\right)^2
-2\sum_{j=1}^M \tilde r_{n,j}|\hat\lambda_j|\\
&=&\frac{1}{n} \sum_{i=1}^n b_i^2\left(\sum_{k=1}^Mu_k\p_k(x_i)\right)^2+2\sum_{j=1}^M \tilde r_{n,j}\left(|\hat\lambda_j+u_j|-|\hat\lambda_j|\right)\\
&&\hspace{1cm}-\frac{2}{n} \sum_{i=1}^n \left(y_i-b_i\sum_{k=1}^M\hat\la_k\p_k(x_i)\right)b_i\sum_{k=1}^Mu_k\p_k(x_i)\\
&=&\frac{1}{n} \sum_{i=1}^n b_i^2\left(\sum_{k=1}^Mu_k\p_k(x_i)\right)^2+2\sum_{j=1}^M \tilde r_{n,j}\left(|\hat\lambda_j+u_j|-|\hat\lambda_j|\right)\\
&&\hspace{1cm}+\frac{2}{n}\sum_{i=1}^n b_i^2\sum_{j=1}^M\hat\la_j\p_j(x_i)\sum_{k=1}^Mu_k\p_k(x_i)-\frac{2}{n}\sum_{i=1}^n b_iy_i\sum_{k=1}^Mu_k\p_k(x_i)\\
&=&\frac{1}{n} \sum_{i=1}^n b_i^2\left(\sum_{k=1}^Mu_k\p_k(x_i)\right)^2+2\sum_{j=1}^M \tilde r_{n,j}\left(|\hat\lambda_j+u_j|-|\hat\lambda_j|\right)\\
&&\hspace{1cm}+\frac{2}{n}\sum_{i=1}^n \sum_{k=1}^Mu_k\p_k(x_i)\left(b_i^2\sum_{j=1}^M\hat\la_j\p_j(x_i)- b_iy_i\right).
\end{eqnarray*}
Since $\hat\la$ minimizes $\la\longmapsto\mbox{crit} (\lambda)$, we have for any $k$,
\begin{eqnarray*}
0
&=&\frac{2}{n}\sum_{i=1}^n \p_k(x_i)\left(b_i^2\sum_{j=1}^M\hat\la_j\p_j(x_i)- b_iy_i\right)+2\tilde r_{n,k}s(\hat\la_k),
\end{eqnarray*}
where $|s(\hat\la_k)|\leq 1$ and $s(\hat\la_k)=\mbox{sign}(\hat\la_k)$ if $\hat\la_k\not=0$. So,
\[\frac{2}{n}\sum_{i=1}^n \sum_{k=1}^Mu_k\p_k(x_i)\left(b_i^2\sum_{j=1}^M\hat\la_j\p_j(x_i)- b_iy_i\right)=-2\sum_{k=1}^Mu_k\tilde r_{n,k}s(\hat\la_k)\]
and
\begin{eqnarray*}
\mbox{crit}(\hat\la+u)- \mbox{crit}(\hat\la)&=&\frac{1}{n} \sum_{i=1}^n b_i^2\left(\sum_{k=1}^Mu_k\p_k(x_i)\right)^2+2\sum_{j=1}^M \tilde r_{n,j}\left(|\hat\lambda_j+u_j|-|\hat\lambda_j|\right)\\
&&\hspace{1cm}-2\sum_{k=1}^Mu_k\tilde r_{n,k}s(\hat\la_k)\\
&=&\frac{1}{n} \sum_{i=1}^n b_i^2\left(\sum_{k=1}^Mu_k\p_k(x_i)\right)^2+2\sum_{j=1}^M \tilde r_{n,j}\left(|\hat\lambda_j+u_j|-|\hat\lambda_j|-u_j s(\hat\la_j)\right)\\
&\geq&\frac{1}{n} \sum_{i=1}^n b_i^2\left(\sum_{k=1}^Mu_k\p_k(x_i)\right)^2,
\end{eqnarray*}
which proves the result.
\hfill$\square$\\\\
Now, still with $s^*=\mbox{card}(S^*)$, we consider for $\mu\in \R^{s^*}$
\begin{equation*}
\mbox{critS}^* (\mu)= \frac{1}{n} \sum_{i=1}^n \left(y_i-b_i \sum_{j\in S^*}\mu_j\p_j(x_i)\right)^2+2\sum_{j\in S^*} \tilde r_{n,j}|\mu_j|,
\end{equation*}
and
\[\tilde\mu=\arg\min_{\mu\in\R^{s^*}}\mbox{critS}^* (\mu).\]
Then we set
\[\mathcal{S}=\bigcap_{j\notin S^*}\left\{ \left|\frac{1}{n}\sum_{i=1}^ny_ib_i \p_j(x_i)-\sum_{k\in S^*}\tilde\mu_k<\p_j,\p_k>\right|< \tilde r_{n,j}\right\}\]
and we state the following lemma.
\begin{lemma}\label{supplemma}
 On the set $\mathcal{S}$, the non-zero coordinates of $\hat\lambda$ are included into $S^*$.
\end{lemma}
{\bf Proof of Lemma~\ref{supplemma}:} Recall that $\hat\lambda$ is a minimizer of $\la\longmapsto\mbox{crit}(\la)$. Using standard convex analysis arguments, this is equivalent to say that for any $1\leq j\leq M$,
\[
\left\{
\begin{array}{lll}
 \frac{1}{n}\sum_{i=1}^ny_ib_i \p_j(x_i)-\sum_{k=1}^M\hat\lambda_k<\p_j,\p_k>&= \tilde r_{n,j}\mbox{sign}(\hat\lambda_j)&\mbox{ if } \hat\lambda_j\not=0,\\
&&\\
\left|\frac{1}{n}\sum_{i=1}^ny_ib_i \p_j(x_i)-\sum_{k=1}^M\hat\lambda_k<\p_j,\p_k>\right|&\leq \tilde r_{n,j}&\mbox{ if } \hat\lambda_j=0.
\end{array}
\right.
\]
Similarly, on $\mathcal{S}$, we have
\[
\left\{
\begin{array}{lll}
 \frac{1}{n}\sum_{i=1}^ny_ib_i \p_j(x_i)-\sum_{k\in S^*}\tilde\mu_k<\p_j,\p_k>&= \tilde r_{n,j}\mbox{sign}(\tilde\mu_j)&\mbox{ if } j\in S^* \mbox{ and } \tilde\mu_j\not=0,\\
&&\\
\left|\frac{1}{n}\sum_{i=1}^ny_ib_i \p_j(x_i)-\sum_{k\in S^*}\tilde\mu_k<\p_j,\p_k>\right|&\leq \tilde r_{n,j}&\mbox{ if } j\in S^* \mbox{ and }\tilde\mu_j=0,\\
&&\\
\left|\frac{1}{n}\sum_{i=1}^ny_ib_i \p_j(x_i)-\sum_{k\in S^*}\tilde\mu_k<\p_j,\p_k>\right|&<\tilde  r_{n,j}&\mbox{ if } j\notin S^*.
\end{array}
\right.
\]
So, on $\mathcal{S}$, the vector $\hat\mu$ such $\hat\mu_j=\tilde\mu_j$ if $j\in S^*$ and $\hat\mu_j=0$ if $j\notin S^*$ is also a minimizer of $\la\longmapsto\mbox{crit}(\la)$. Using Lemma~\ref{unicite}, we have for any $1\leq i\leq n$:
\[\sum_{k=1}^M(\hat\lambda_k-\hat\mu_k)\p_k(x_i)=0.\]
So, for $j\notin S^*$,
\[\left|\frac{1}{n}\sum_{i=1}^ny_ib_i \p_j(x_i)-\sum_{k=1}^M\hat\lambda_k<\p_j,\p_k>\right|< \tilde r_{n,j}.\] Therefore, on $\mathcal{S}$,
the non-zero coordinates of $\hat\lambda$ are included into $S^*$.\hfill$\square$\\\\
Lemma~\ref{supplemma} shows that we just need to prove that \[\P\left\{ \mathcal{S}\right\}\geq 1-2M^{1-\gamma/2}\]
\begin{eqnarray*}
\P\left\{ \mathcal{S}^c\right\}&\leq&\sum_{j\notin S^*}\P\left\{\left|\frac{1}{n}\sum_{i=1}^ny_ib_i \p_j(x_i)-\sum_{k\in S^*}\tilde\mu_k<\p_j,\p_k>\right|\geq \tilde r_{n,j}\right\}\\
&\leq&A+B,
\end{eqnarray*}
with
\begin{eqnarray*}
A&=&\sum_{j\notin S^*}\P\left\{\left|\frac{1}{n}\sum_{i=1}^n\left[y_ib_i \p_j(x_i)-\E(y_ib_i \p_j(x_i))\right]\right|\geq r_{n,j}\right\}\\
&=&\sum_{j\notin S^*}\P\left\{\left|\frac{1}{n}\sum_{i=1}^n\e_i b_i \p_j(x_i)\right|\geq r_{n,j}\right\}\\
&=&\sum_{j\notin S^*}\P\left\{\left| V_j\right|\geq r_{n,j}\right\}
\end{eqnarray*}
(see Lemma~\ref{prel}) and
\begin{eqnarray*}
B&=&\P\left[\bigcup_{j\notin S^*}\left\{\left|\frac{1}{n}\sum_{i=1}^n\E(y_ib_i \p_j(x_i))
-\sum_{k\in S^*}\tilde\mu_k<\p_j,\p_k>\right|\geq \tilde r_{n,j}-r_{n,j}\right\}\right]\\
&=&\P\left[\bigcup_{j\notin S^*}\left\{\left|<\p_j,f_{\la^*}>
-\sum_{k\in S^*}\tilde\mu_k<\p_j,\p_k>\right|\geq \tilde r_{n,j}-r_{n,j}\right\}\right]\\
&=&\P\left[\bigcup_{j\notin S^*}\left\{\left|\sum_{k\in S^*}(\la^*_k-\tilde\mu_k)<\p_j,\p_k>\right|\geq \tilde r_{n,j}-r_{n,j}\right\}\right]\\
&\leq&\P\left[\bigcup_{j\notin S^*}\left\{\rho(S^*)\|\p_j\|_n\sum_{k\in S^*}\left|\la^*_k-\tilde\mu_k\right|\|\p_k\|_n\geq \tilde r_{n,j}-r_{n,j}\right\}\right]
\end{eqnarray*}
since
\[\rho(S^*)=\max_{k\in S^*}\max_{ j\not=k}\frac{|<\p_j,\p_k>|}{\|\p_j\|_n\|\p_k\|_n}.\]
Using notation of Lemma~\ref{supplemma}, we have:
\begin{eqnarray*}
\|f_{\la^*}-f_{\hat\mu}\|_n^2&=&\|\sum_{k\in S^*}(\la^*_k-\hat\mu_k)\p_k\|_n^2\\
&=&\sum_{k\in S^*}(\la^*_k-\hat\mu_k)^2\|\p_k\|_n^2+\sum_{k\in S^*}\sum_{j\in S^*,\ j\not=k}(\la^*_k-\hat\mu_k)(\la^*_j-\hat\mu_j)<\p_j,\p_k>,
\end{eqnarray*}
and
\begin{eqnarray*}
\sum_{k\in S^*}(\la^*_k-\hat\mu_k)^2\|\p_k\|_n^2&\leq&\|f_{\la^*}-f_{\hat\mu}\|_n^2+\rho(S^*)\sum_{k\in S^*}\sum_{j\in S^*, \ j\not=k}|\la^*_k-\hat\mu_k|\|\p_k\|_n\times|\la^*_j-\hat\mu_j|\|\p_j\|_n\\
&\leq&\|f_{\la^*}-f_{\hat\mu}\|_n^2+\rho(S^*)\left(\sum_{k\in S^*}|\la^*_k-\hat\mu_k|\|\p_k\|_n\right)^2.
\end{eqnarray*}
Finally,
\begin{eqnarray*}
\left(\sum_{k\in S^*}|\la^*_k-\hat\mu_k|\|\p_k\|_n\right)^2&\leq&s^*\sum_{k\in S^*}(\la^*_k-\hat\mu_k)^2\|\p_k\|_n^2\\
&\leq&s^*\left(\|f_{\la^*}-f_{\hat\mu}\|_n^2+\rho(S^*)\left(\sum_{k\in S^*}|\la^*_k-\hat\mu_k|\|\p_k\|_n\right)^2\right),
\end{eqnarray*}
which shows that
\[\left(\sum_{k\in S^*}|\la^*_k-\hat\mu_k|\|\p_k\|_n\right)^2\leq\frac{s^*}{1-\rho(S^*)s^*}\|f_{\la^*}-f_{\hat\mu}\|_n^2.\]
Now,
\begin{eqnarray*}
&&\frac{1}{n} \sum_{i=1}^n \left(y_i-b_i \sum_{j\in S^*}\tilde\mu_j\p_j(x_i)\right)^2+2\sum_{j\in S^*} \tilde r_{n,j}|\tilde\mu_j|\leq\\
&&\hspace{2cm}
\frac{1}{n} \sum_{i=1}^n \left(y_i-b_i \sum_{j\in S^*}\la_j^*\p_j(x_i)\right)^2+2\sum_{j\in S^*} \tilde r_{n,j}|\la_j^*|.
\end{eqnarray*}
So,
\begin{eqnarray*}
&&\| \sum_{j\in S^*}\tilde\mu_j\p_j\|_n^2-\frac{2}{n}\sum_{i=1}^nb_iy_i\sum_{j\in S^*}\tilde\mu_j\p_j(x_i)+2\sum_{j\in S^*} \tilde r_{n,j}|\tilde\mu_j|\leq\\
&&\hspace{2cm}\| \sum_{j\in S^*}\la_j^*\p_j\|_n^2-\frac{2}{n}\sum_{i=1}^nb_iy_i\sum_{j\in S^*}\la_j^*\p_j(x_i)+2\sum_{j\in S^*} \tilde r_{n,j}|\la_j^*|,
\end{eqnarray*}
and using previous notation,
 \begin{eqnarray*}
&&\| f_{\hat\mu}\|_n^2-\frac{2}{n}\sum_{i=1}^nb_iy_i\sum_{j\in S^*}\tilde\mu_j\p_j(x_i)+2\sum_{j\in S^*} \tilde r_{n,j}|\tilde\mu_j|\leq\\
&&\hspace{2cm}\|  f_{\la^*}\|_n^2-\frac{2}{n}\sum_{i=1}^nb_iy_i\sum_{j\in S^*}\la_j^*\p_j(x_i)+2\sum_{j\in S^*} \tilde r_{n,j}|\la_j^*|.
\end{eqnarray*}
Therefore,
\begin{eqnarray*}
\|f_{\la^*}-f_{\hat\mu}\|_n^2&=&\|f_{\hat\mu}\|_n^2+\|f_{\la^*}\|_n^2-2<f_{\hat\mu},f_{\la^*}>\\
&\leq&2\|f_{\la^*}\|_n^2-2<f_{\hat\mu},f_{\la^*}>+\frac{2}{n}\sum_{i=1}^nb_iy_i\sum_{j\in S^*}(\tilde\mu_j-\la_j^*)\p_j(x_i)+2\sum_{j\in S^*} \tilde r_{n,j}(|\la_j^*|-|\tilde\mu_j|)\\
&=&\frac{2}{n}\sum_{i=1}^nb_iy_i(f_{\hat\mu}(x_i)-f_{\la^*}(x_i))-\frac{2}{n}\sum_{i=1}^nb_i^2f_{\la^*}(x_i)(f_{\hat\mu}(x_i)-f_{\la^*}(x_i))\\&&\hspace{2cm}+2\sum_{j\in S^*} \tilde r_{n,j}(|\la_j^*|-|\tilde\mu_j|)\\
&=&\frac{2}{n}\sum_{i=1}^nb_i(y_i-\E(y_i))(f_{\hat\mu}(x_i)-f_{\la^*}(x_i))+2\sum_{j\in S^*} \tilde r_{n,j}(|\la_j^*|-|\tilde\mu_j|)\\
&=&\frac{2}{n}\sum_{i=1}^nb_i\e_i(f_{\hat\mu}(x_i)-f_{\la^*}(x_i))+2\sum_{j\in S^*} \tilde r_{n,j}(|\la_j^*|-|\tilde\mu_j|)\\
&=&2\sum_{j=1}^MV_j(\hat\mu_j-\la^*_j)+2\sum_{j\in S^*} \tilde r_{n,j}(|\la_j^*|-|\tilde\mu_j|).
\end{eqnarray*}
Now let us assume that for any $j\in S^*$, $V_j<r_{n,j}.$ Then,
\begin{eqnarray*}
\|f_{\la^*}-f_{\hat\mu}\|_n^2&<& 2\sum_{j\in S^*} (r_{n,j}+\tilde r_{n,j})|\hat\mu_j-\la^*_j|\\
&<&2\sigma\sqrt{\frac{\log M}{n}}(\sqrt{\gamma}+\sqrt{\tilde\gamma})\sum_{j\in S^*}\|\p_j\|_n|\hat\mu_j-\la^*_j|.
\end{eqnarray*}
So,
\begin{eqnarray*}
\sum_{k\in S^*}|\la^*_k-\hat\mu_k|\|\p_k\|_n&<& 2\sigma\sqrt{\frac{\log M}{n}}(\sqrt{\gamma}+\sqrt{\tilde\gamma})\frac{s^*}{1-\rho(S^*)s^*}
\end{eqnarray*}
and for any $j\notin S^*$,
\begin{eqnarray*}
\rho(S^*)\|\p_j\|_n\sum_{k\in S^*}|\la^*_k-\hat\mu_k|\|\p_k\|_n&<& 2\sigma\sqrt{\frac{\log M}{n}}\|\p_j\|_n(\sqrt{\gamma}+\sqrt{\tilde\gamma})\frac{\rho(S^*)s^*}{1-\rho(S^*)s^*}\\
&<&\frac{2\sigma c(\sqrt{\gamma}+\sqrt{\tilde\gamma})}{1-c}\sqrt{\frac{\log M}{n}}\|\p_j\|_n\\
&<&(\sqrt{\tilde\gamma}-\sqrt{\gamma})\sigma\sqrt{\frac{\log M}{n}}\|\p_j\|_n\\
&<&\tilde r_{n,j}-r_{n,j}.
\end{eqnarray*}
Therefore,
\begin{eqnarray*}
B&\leq&\sum_{j\in S^*}\P\left\{\left| V_j\right|\geq r_{n,j}\right\}
\end{eqnarray*}
and using Lemma~\ref{prel}, since $\P\left\{ \mathcal{S}^c\right\}\leq A+B$,
\[\P\left\{ \mathcal{S}\right\}\geq 1-2M^{1-\gamma/2}.\]

\subsection*{Proof of Corollary~\ref{oracleadmi3}}

First note that $\lambda^*$ satisfies the Dantzig constraint (\ref{dantzig}) where $r_{n,j}$ is replaced by $\tilde r_{n,j}$ with probability larger than $1-M^{1-\tilde{\gamma}/2}$.
On the event $\hat{S}\subset S^*$, we have $\lambda^*_{(S^*)^C}=\hat{\lambda}_{(S^*)^C}=0$, then applying Theorem~\ref{oracleadmi2}, we obtain that for any $\alpha>0$
\[\normD{\hat{f}-\fo}^2 \leq  32s^*
\left(\frac{1}{\betamod}+\frac{1}{\kappa_{s^*}^2}\right)
\tilde{r}_n^2,
\]
which implies the result of the theorem.

\bibliographystyle{apalike}
\bibliography{biblio_ncspnlmem}

\begin{thebibliography}{}

\bibitem[Bertin et~al., 2011]{berlepenrivo}
Bertin, K., Le~Pennec, E., and Rivoirard, V. (2011).
\newblock Adaptive {D}antzig density estimation.
\newblock {\em Annales de l'Institut Henri Poincaré}, 47:43--74.

\bibitem[Bickel et~al., 2009]{BRT}
Bickel, P.~J., Ritov, Y., and Tsybakov, A.~B. (2009).
\newblock Simultaneous analysis of lasso and {D}antzig selector.
\newblock {\em Ann. Statist.}, 37(4):1705--1732.

\bibitem[B{\"u}hlmann and van~de Geer, 2011]{livreBulmannSara}
B{\"u}hlmann, P. and van~de Geer, S. (2011).
\newblock {\em Statistics for high-dimensional data}.
\newblock Springer Series in Statistics. Springer, Heidelberg.
\newblock Methods, theory and applications.

\bibitem[Bunea, 2008]{bunea2}
Bunea, F. (2008).
\newblock Consistent selection via the {L}asso for high dimensional
  approximating regression models.
\newblock In {\em Pushing the limits of contemporary statistics: contributions
  in honor of {J}ayanta {K}. {G}hosh}, volume~3 of {\em Inst. Math. Stat.
  Collect.}, pages 122--137. Inst. Math. Statist., Beachwood, OH.

\bibitem[Bunea et~al., 2007a]{btw}
Bunea, F., Tsybakov, A., and Wegkamp, M. (2007a).
\newblock Sparsity oracle inequalities for the {L}asso.
\newblock {\em Electronic Journal of Statistics}, 1:169--194.

\bibitem[Bunea et~al., 2006]{btw06}
Bunea, F., Tsybakov, A.~B., and Wegkamp, M.~H. (2006).
\newblock Aggregation and sparsity via {$l_1$} penalized least squares.
\newblock In {\em Learning theory}, volume 4005 of {\em Lecture Notes in
  Comput. Sci.}, pages 379--391. Springer, Berlin.

\bibitem[Bunea et~al., 2007b]{Gau}
Bunea, F., Tsybakov, A.~B., and Wegkamp, M.~H. (2007b).
\newblock Aggregation for {G}aussian regression.
\newblock {\em The Annals of Statistics}, 35(4):1674--1697.

\bibitem[Cand\`es and Tao, 2007]{candes}
Cand\`es, E.~J. and Tao, T. (2007).
\newblock The {D}antzig selector: statistical estimation when $p$ is much
  larger than $n$.
\newblock {\em The Annals of Statistics}, 35(6):2313--2351.

\bibitem[Comte and Samson, 2012]{Adeline}
Comte, F. and Samson, A. (2012).
\newblock Nonparametric estimation of random effects densities in linear
  mixed-effects model.
\newblock Unpublished manuscript. Available at
  http://hal.archives-ouvertes.fr/hal-00657052/fr/.

\bibitem[Delyon et~al., 1999]{saem}
Delyon, B., Lavielle, M., and Moulines, E. (1999).
\newblock Convergence of a stochastic approximation version of the em
  algorithm.
\newblock {\em The Annals of Statistics}, 27:94--128.

\bibitem[Dempster et~al., 1977]{EM}
Dempster, A.~P., Laird, N.~M., and Rubin, D.~B. (1977).
\newblock Maximum-likelihood from incomplete data via the {EM} algorithm.
\newblock {\em Journal of Royal Statistical Society, Series B}, 39:1--38.

\bibitem[Ding and Wu, 2001]{DingWu}
Ding, A.~A. and Wu, H. (2001).
\newblock Assessing antiviral potency of anti-{HIV} therapies in vivo by
  comparing viral decay rates in viral dynamic models.
\newblock {\em Biostatistics}, 2:13--29.

\bibitem[Hartford and Davidian, 2000]{nlme_pb}
Hartford, A. and Davidian, M. (2000).
\newblock Consequences of misspecifying assumptions in nonlinear mixed effects
  models.
\newblock {\em Computational Statistics \& Data Analysis}, 34:139--164.

\bibitem[Harville, 1974]{REML2}
Harville, D. (1974).
\newblock Bayesian inference for variance components using only error
  contrasts.
\newblock {\em Biometrika}, 61:383--385.

\bibitem[Jank and Shmueli, 2006]{AD2}
Jank, W. and Shmueli, G. (2006).
\newblock Functional data analysis in electronic commerce research.
\newblock {\em Statistical Science}, 21:155--166.

\bibitem[Ke and Wang, 2001]{KeWang}
Ke, C. and Wang, Y. (2001).
\newblock Semiparametric nonlinear mixed-effects models and their applications
  (with discussion).
\newblock {\em Journal of the American Statistical Association},
  96(456):1272--1298.

\bibitem[Ke and Wang, 2004]{KeWang04}
Ke, C. and Wang, Y. (2004).
\newblock Smoothing spline nonlinear nonparametric regression models.
\newblock {\em Journal of the American Statistical Association},
  99(468):1166--1175.

\bibitem[Kuhn and Lavielle, 2004]{Kuhn04}
Kuhn, E. and Lavielle, M. (2004).
\newblock Coupling a stochastic approximation version of {EM} with an {MCMC}
  procedure.
\newblock {\em ESAIM: P\&S}, 8:115--131.

\bibitem[Kuhn and Lavielle, 2005]{Kuhn05}
Kuhn, E. and Lavielle, M. (2005).
\newblock Maximum likelihood estimation in nonlinear mixed effects models.
\newblock {\em Computational Statistics {\&} Data Analysis}, 49(4):1020--1038.

\bibitem[Liu and M{\"{u}}ller, 2008]{AD4}
Liu, B. and M{\"{u}}ller, H.~G. (2008).
\newblock Functional data analysis for sparse auction data.
\newblock In Jank, W. and Shmueli, G., editors, {\em Statistical Methods in
  {E}-commerce research}, pages 269--290. Wiley, New York.

\bibitem[Liu and Wu, 2007]{LiuWu07}
Liu, W. and Wu, L. (2007).
\newblock Simultaneous inference for semiparametric nonlinear mixed-effects
  models with covariate measurement errors and missing responses.
\newblock {\em Biometrics}, 63:342--350.

\bibitem[Liu and Wu, 2008]{LiuWu08}
Liu, W. and Wu, L. (2008).
\newblock A semiparametric nonlinear mixed-effects model with non-ignorable
  missing data and measurement errors for {HIV} viral data.
\newblock {\em Computational Statistics \& Data Analysis}, 53:112--122.

\bibitem[Liu and Wu, 2009]{LiuWu09}
Liu, W. and Wu, L. (2009).
\newblock Some asymptotic results for semiparametric nonlinear mixed-effects
  models with incomplete data.
\newblock {\em Journal of Statistical Planning and Inference}.
\newblock doi:10.1016\/j.jspi.2009.06.006.

\bibitem[Luan and Li, 2004]{LuanLi}
Luan, Y. and Li, H. (2004).
\newblock Model-based methods for identifying periodically expressed genes
  based on time course microarray gene expression data.
\newblock {\em Bioinformatics}, 20(3):332--339.

\bibitem[Meza et~al., 2007]{Cristian1}
Meza, C., Jaffrézic, F., and Foulley, J.-L. (2007).
\newblock Reml estimation of variance parameters in nonlinear mixed effects
  models using the {SAEM} algorithm.
\newblock {\em Biometrical Journal}, 49(6):876--888.

\bibitem[Patterson and Thompson, 1971]{REML1}
Patterson, H.~D. and Thompson, R. (1971).
\newblock Recovery of inter-block information when block sizes are unequal.
\newblock {\em Biometrika}, 58:545--554.

\bibitem[Pinheiro and Bates, 2000]{Pin00}
Pinheiro, J. and Bates, D. (2000).
\newblock {\em Mixed-Effects Models in {S} and {S-PLUS}}.
\newblock Springer-Verlag, New York.

\bibitem[Ramos and Pantula, 1995]{RamosPantula}
Ramos, R. and Pantula, S. (1995).
\newblock Estimation of nonlinear random coefficient models.
\newblock {\em Statistics \& Probability Letters}, 24:49--56.

\bibitem[Reithinger et~al., 2008]{AD_mm}
Reithinger, F., Jank, W., Tutz, G., and Shmueli, G. (2008).
\newblock Modelling price paths in on-line auctions: smoothing sparse and
  unevenly sampled curves by using semiparametric mixed models.
\newblock {\em Applied Statistics}, 57:127--148.

\bibitem[Schelldorfer et~al., 2011]{vdgeer}
Schelldorfer, J., Bühlmann, P., and van~de Geer, S. (2011).
\newblock Estimation for high-dimensional linear mixed-effects models using
  l1-penalization.
\newblock {\em Scandinavian Journal of Statistics}, 38:197--214.

\bibitem[Shmueli and Jank, 2005]{AD1}
Shmueli, G. and Jank, W. (2005).
\newblock Visualizing online auctions.
\newblock {\em Journal of Computational and Graphical Statistics}, 14:299--319.

\bibitem[Shmueli et~al., 2007]{AD3}
Shmueli, G., Russo, R.~P., and Jank, W. (2007).
\newblock The {BARISTA}: a model for bid arrivals in online auctions.
\newblock {\em The Annals of Applied Statistics}, 1:412--441.

\bibitem[Tibshirani, 1996]{LASSO}
Tibshirani, R. (1996).
\newblock Regression shrinkage and selection via the {L}asso.
\newblock {\em Journal of the Royal Statistical Society, Series B},
  58:267--288.

\bibitem[van~de Geer, 2010]{HighSara}
van~de Geer, S. (2010).
\newblock {$\ell_1$}-regularization in high-dimensional statistical models.
\newblock In {\em Proceedings of the {I}nternational {C}ongress of
  {M}athematicians. {V}olume {IV}}, pages 2351--2369, New Delhi. Hindustan Book
  Agency.

\bibitem[van~de Geer and B{\"u}hlmann, 2009]{LassoAss}
van~de Geer, S.~A. and B{\"u}hlmann, P. (2009).
\newblock On the conditions used to prove oracle results for the {L}asso.
\newblock {\em Electronic Journal of Statistics}, 3:1360--1392.

\bibitem[Vonesh, 1996]{Vonesh96}
Vonesh, E.~F. (1996).
\newblock A note on the use of {L}aplace's approximation for nonlinear
  mixed-effects models.
\newblock {\em Biometrika}, 83:447--452.

\bibitem[Wahba, 1990]{Wahba}
Wahba, G. (1990).
\newblock {\em Spline models for observational data}, volume~59 of {\em
  CBMS-NSF Regional Conference Series in Applied Mathematics}.
\newblock Society for Industrial and Applied Mathematics (SIAM), Philadelphia,
  PA.

\bibitem[Wang, 1998]{Wang98}
Wang, Y. (1998).
\newblock Smoothing spline models with correlated random errors.
\newblock {\em Journal of the American Statistical Association}, 93:341--348.

\bibitem[Wang and Brown, 1996]{WangBrown}
Wang, Y. and Brown, M.~B. (1996).
\newblock A flexible model for human circadian rhythms.
\newblock {\em Biometrics}, 52:588--596.

\bibitem[Wang et~al., 2003]{WangKeBrown}
Wang, Y., Ke, C., and Brown, M.~B. (2003).
\newblock Shape-invariant modeling of circadian rhythms with random effects and
  smoothing spline anova decompositions.
\newblock {\em Biometrics}, 59:804--812.

\bibitem[Wei and Tanner, 1990]{Wei90}
Wei, G.~C. and Tanner, M.~A. (1990).
\newblock A {M}onte {C}arlo implementation of the {EM} algorithm and the poor
  man's data augmentation algorithm.
\newblock {\em Journal of the American Statistical Association}, 85:699--704.

\bibitem[Wu and Zhang, 2002]{WuZhang}
Wu, H. and Zhang, J. (2002).
\newblock The study of longterm {HIV} dynamics using semi-parametric non-linear
  mixed-effects models.
\newblock {\em Statistics in Medicine}, 21:3655--3675.

\bibitem[Yuan and Lin, 2006]{YL06}
Yuan, M. and Lin, Y. (2006).
\newblock Model selection and estimation in regression with grouped variables.
\newblock {\em Journal of the Royal Statistical Society, Series B},
  68(1):49--67.

\end{thebibliography}

\end{document}